\newif\ifdraft
\drafttrue

\documentclass[USenglish,oneside,twocolumn]{article}

\usepackage[utf8]{inputenc}
\usepackage[big]{dgruyter_NEW}

\usepackage[usenames,dvipsnames]{xcolor}
\usepackage{amssymb,amsfonts,euscript}
\usepackage{xspace}
\usepackage{hyperref}
\usepackage[capitalize]{cleveref}
\usepackage{paralist}
\usepackage{times,framed}
\usepackage{balance}
\usepackage[noend]{algpseudocode}
\usepackage{algorithm}
\usepackage{verbatim}
\usepackage{subcaption}
\usepackage{makecell}
\usepackage[scr]{rsfso}

 \newtheorem{definition}{Definition}
\newtheorem{theorem}{Theorem}

\newcommand{\myOT}[2]{\textrm{OT}^{#1}_{#2}}
\newcommand{\mysavespace}{\vspace*{-8pt}}
\newcommand{\mysavemorespace}{\vspace*{-12pt}}
\newcommand{\mysavelessspace}{\vspace*{-6pt}}
\newcommand{\mysavetinyspace}{\vspace*{-2pt}}

\begin{document}
 

  \author*[1]{Farid Javani}

  \author[2]{Alan T. Sherman}



  \affil[1]{Cyber Defense Lab, University of Maryland, Baltimore County, USA, E-mail: javani1@umbc.edu}

  \affil[2]{Cyber Defense Lab, University of Maryland, Baltimore County, USA, E-mail: sherman@umbc.edu}



  \title{\huge AOT: Anonymization by Oblivious Transfer}

  \runningtitle{AOT: Anonymization by Oblivious Transfer}


  \begin{abstract}
{We introduce AOT, an anonymous communication system based on mix network architecture that uses {\it oblivious transfer (OT)} to deliver messages.
Using OT to deliver messages helps AOT resist blending ($n-1$) attacks and 
helps AOT preserve receiver anonymity, even if a covert adversary controls all nodes in AOT.\newline
AOT comprises three levels of nodes, where nodes at each level perform a different function and can scale horizontally. The sender encrypts their payload and a tag---derived from a secret shared between the sender and receiver---with the public key of a Level-2 node and sends them to a Level-1 node. On a public bulletin board, Level-3 nodes publish tags associated with messages ready to be retrieved.
Each receiver checks the bulletin board, identifies tags, and receives the associated messages using OT. \newline
A receiver can receive their messages even if the receiver is offline when messages are ready. 
Through what we call a ``handshake'' process, communicants can use the AOT protocol to establish shared secrets anonymously.\newline
Users play an active role in contributing to the unlinkability of messages:
periodically, users initiate requests to AOT to receive dummy messages,
such that an adversary cannot distinguish real and dummy requests.}
\end{abstract}


\maketitle

\mysavetinyspace
\section{Introduction}

Communication systems should not only support confidentiality and authentication of messages; they should also provide untraceability and unlinkability. Failure to protect communications against traffic analysis poses serious threats to individual privacy and organizational operations.
We introduce AOT~\cite{FJredacted}, 
an {\it anonymous communication system (ACS)} 
based on {\it mix network (mixnet)} architecture
that uses {\it oblivious transfer (OT)} to deliver messages.  Our approach
resists active attacks, supports message delivery for offline users, and provides receiver anonymity even if a covert active adversary controls the entire network.


Although Kilian~\cite{kil88} proved that OT is complete for two-party secure computations, to our knowledge, we are the first to show how to build an ACS with OT.  Desirable properties of AOT mentioned above flow in part from the use of OT.

\begin{figure}[ht!] 
\centering
	\includegraphics[width=.85\columnwidth]{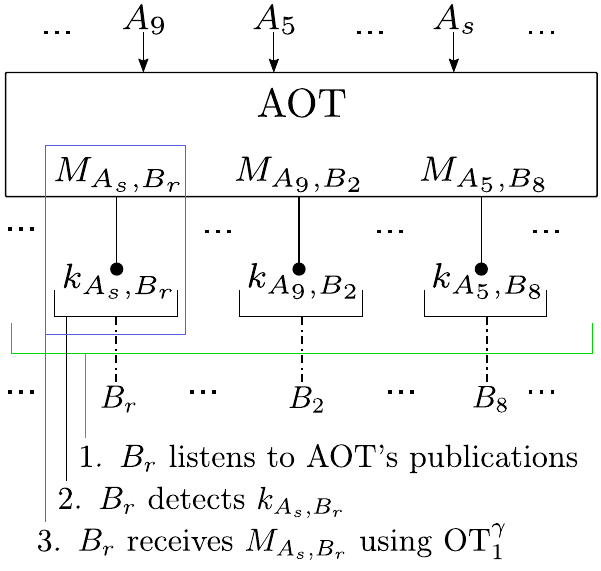}
	\caption{Message delivery using OT. 
	When sender $A_s$ sends a message through AOT to $B_r$, receiver $B_r$ retrieves the message
	as follows. $A_s$ sends an ordered pair $(M_{A_s,B_r}, k_{A_s,B_r})$ to AOT, where 
	$M_{A_s,B_r}$ is the encrypted payload and $k_{A_s,B_r}$ a tag derived from a  
	shared secret between $A_s$ and $B_r$.  The payload includes the message and some
	additional information, as described in Equation~\ref{eq:format}. 
	AOT publishes $k_{A_t,B_t}$. Then,
	$B_t$ detects $k_{A_t,B_t}$ and asks for the corresponding encrypted payload 
	using $\myOT{\gamma}{1}$, where $\gamma$ is the number 
	of other tags that $B_t$ chooses to be in the OT session.
	Only $B_r$ can decrypt the encrypted payload, and AOT does not learn which message
	$B_r$ received.}
	\label{fig:OT}
\end{figure}

Figure~\ref{fig:OT} highlights the keystone of AOT---message delivery using OT. In AOT, a sender does not provide the recipient's address.  Instead, they send an ordered pair $(M, \hbox{tag})$, where $M$ is an encrypted payload (encrypted with the public key of the recipient), and {\it tag} is a unique tag derived from a {\it shared secret} between  sender and receiver.  For messages ready to be retrieved, AOT posts their associated tags on a public bulletin board.  Recipients monitor the board and request via OT the payloads they wish to receive.  Neither AOT nor network adversaries learn which recipients receive which messages.

{Instead of using OT to retrieve messages, one could use
{\it private information retrieval (PIR)}~\cite{chor95pir}. 
Whereas PIR provides only receiver security, OT provides both
receiver and sender security 
(see Sections~\ref{sec:RACovert}, \ref{sec:decisions}).
Using OT helps AOT resist active attacks.}

AOT is a mixnet comprising a three-level cascade of nodes, where each level performs a different function. The sender encrypts their payload and tag
with the public key of a Level-2 node and sends them to a Level-1 node. 
Level-1 nodes strip the sender information from messages and send them to Level-2 nodes in batches. Level-2 nodes decrypt the messages, create dummy messages, and send the real and the dummy messages to Level-3 nodes in batches. 
Dummy messages help resist blending ($n-1$) attacks.
At each level, all nodes at that level perform the same function 
and can scale horizontally (more nodes can be added at each level to improve performance and reliability).


Each sender-receiver pair needs to share a {secret}.
Through what we call a ``handshake'' process, communicants can use the AOT protocol to establish shared secrets, confidentially and anonymously.  This handshake is of independent interest and
can support other applications.

Since Chaum~\cite{chaum81} introduced ACSs in 1981, they have 
evolved in terms of efficiency, network topology, communication latency, robustness,
and privacy~\cite{jakobsson2001,park93,tor2004,kwon2016}. 
For example, cMix by Chaum et al.~\cite{chaum2017cmix, cMix-patent} uses precomputation to support
fast mixing with minimal real-time asymmetric cryptographic operations.
Many recent systems,
including~\cite{wolinsky2012dissent,van2015vuvuzela,tyagi2017stadium,piotrowska2017loopix,kwon2016},
aim to resist active attacks, as does AOT.
{An advantage of AOT over cMix is its resistance to active attacks.}

Our contributions:

\begin{enumerate}

    \item We introduce AOT, an anonymous communication system that uses oblivious transfer for message delivery and to resist active attacks.
    
    \item AOT scales horizontally in that additional nodes can be added to each level to increase performance and reliability.
    
    \item We show how two communicants can use AOT to establish a shared secret anonymously.

    \item We explain how AOT resists standard active and passive mixnet attacks.
    
\end{enumerate}


\mysavetinyspace
\section{Background and Related work}
\label{sec:background}

We briefly overview {\it anonymous communication systems (ACSs)}, 
{\it oblivious transfer (OT)}, and 
applications of OT in selected cryptographic protocols.
To our knowledge, we are the first to use OT to build an ACS.

\mysavespace
\subsection{Anonymous Communication Networks}

In 1981, Chaum~\cite{chaum81} introduced the concept of {\it mix networks}, or {\it mixnets} for short. Mixnets are ACSs, where a cascade of servers, called {\it mixnodes}, break the links between communicants. Each sender encrypts the payload and the address of the receiver using the public keys of the mixnodes, starting from the last node in the cascade, and then sends the message to the first node in the cascade. Upon receiving an encrypted message, each mixnode decrypts the outer layer of encryption using its private key. Each mixnode gathers incoming messages into a fixed-size batch, shuffles the messages within the batch, and sends the batch to the next node in the cascade. The last node in the cascade decrypts the last layer of the encryption and delivers the messages to the recipients.

To prevent an adversary from linking incoming and outgoing messages based on length, mixnets typically allow only messages of a fixed length. \textit{Hybrid mixnets}, introduced by Pfitzmann and Waidner~\cite{pfitzmann85}, allow arbitrary length messages. Hybrid mixnets perform bulk data encryption with symmetric-key cryptography; 
they use public-key cryptography to encrypt symmetric keys.  
Jackobson et al.~\cite{jakobsson2001}, 
Ohkubo et al.~\cite{ohkubo2000l}, and Moller~\cite{moller2003} 
propose combinations of public-key cryptography and symmetric-key cryptography 
for hybrid mixnets.

In mixnets and hybrid mixnets, message length is proportional to the number of mixnodes, because the sender has to perform one encryption for each node in the cascade. These layers of encryption increase computation
on the user side and increase message length.
In \textit{re-encryption mixnets}, introduced by Park {et al.}~\cite{park93}, the sender encrypts only once using the public key of the mixnet. 
Instead of decrypting the outer later of encryption, each mixnode
re-encrypts the ciphertext received from the previous node (using an encryption system that permits re-encryption) and forwards it to the next node. 
The last node in the cascade produces plaintext.
Golle et al.~\cite{golle04} introduce universal re-encryption mixnets, where re-encrytion does not require knowledge of the public key.

Most mixnets rely heavily on users and servers to perform computationally expensive public-key operations online. 
Performing such operations in an offline precomputation can significantly
improve the online running time.
Adida and Wikstrom~\cite{AdidaW07} and Jakobsson~\cite{jakobsson1999flash}
study precomputation techniques, with
Adida and Wikstrom focusing on server computations,
and Jakobsson focusing on user computations.
Chaum {et al.}'s {\it cMix}~\cite{chaum2017cmix}
is a precomputation mixnet that eliminates almost all online public-key operations.

By routing messages through a cascade of nodes, mixnets increase 
message latency.  They also can introduce delays caused by grouping messages
into the batches needed for anonymization. Therefore, mixnets, despite their strong anonymity properties, are not suitable for low-latency applications such as web browsing.

\textit{Onion routing}~\cite{reed98} is an anonymization technique that 
neither uses batches nor a fixed cascade of nodes.  In comparison with mixnets, they can have lower latency but often are vulnerable to traffic analysis
attacks~\cite{rackoff1993cryptographic,raymond2001traffic}.
Intermediary proxy nodes break the communication links, and they can vary with each communication session.
Each sender may choose the communication path and which nodes will
act as the proxy routers.
The sender encrypts a message using the keys of the proxies in the path, starting with the key of the last node. 
Proxies decrypt the (onion) layers of encryption and forward the message. 
{\it Tor}~\cite{tor2004} is a widely used system built using onion routing; 
other examples of ACSs built on onion routing include \cite{camenisch2005,overlier2007, chen2015}.

{\it Riffle}~\cite{kwon2016} is an anonymous communication and file sharing system in which users download messages or files
using {\it private information retrieval (PIR)}~\cite{chor95pir}.
Riffle focuses on communications between users within a group. 
To enhance anonymity, all users should send and receive messages even if they are not participating in any communications. Messages are initially broadcast to all group members. After users learn the index of the message corresponding to them within the group, they can receive messages using this index and PIR---instead of receiving all messages through broadcast.

{\it Dissent}~\cite{wolinsky2012dissent} is a group ACS 
build on DC-nets~\cite{chaum1988dining} and verifiable shuffles~\cite{brickell2006efficient, furukawa2001efficient}. Leveraging a user-server architecture, Dissent increases the efficiency of DC-nets and tolerates user-side slowness and disruptions.

{\it Loopix}~\cite{piotrowska2017loopix} is a mixnet in which nodes are grouped in different layers, where nodes in each layer can communicate with all of the nodes in the immediately previous and following layers. 
Loopix adds independent delays to incoming message (Poisson mixing) to obfuscate message timing.
An important property of Loopix is the so called ``loop message,'' 
a message that a sender sends to itself; 
loop messages help Loopix resist blending attacks (see Section \ref{sec:blending}). {As explained in Section~\ref{sec:clnt}, every message in AOT is a loop message.}

{\it Vuvuzela}~\cite{van2015vuvuzela} is an efficient anonymous messaging protocol that can support millions of users. It is built on a mixnet architecture and offers privacy guarantees based on differential privacy~\cite{dwork2014algorithmic}.
Users place and retrieve messages from virtual locations (``deaddrops'') on mixnode memories. Vuvuzula has a ``dialing protocol'' (similar to AOT's handshake) that users can use to start a conversation.
Unlike AOT's handshake, Vuvuzela's dialing protocol is distinct from its conversation protocol. Moreover, in Vuvuzela, the adversary can distinguish whether a receiver is participating in a dialing or conversation protocol, whereas in AOT the adversary cannot make this distinction.
{\it Stadium}~\cite{tyagi2017stadium} adds horizontal scalability and verifiable mixing to Vuvuzela. Deaddrops and AOT's message tags (See Section~\ref{sec:ud})
are similar to the rendezvous hash value of the UDM protocol of Chaum et al.~\cite{CHaumUDM}.

\mysavespace
\subsection{Oblivious Transfer}

We briefly review our main building block---{\it oblivious transfer (OT)}---including selected OT protocols and their security and efficiency.

Introduced in 1981 by Rabin~\cite{Rabin81}, an OT protocol enables a receiver to receive a piece of information from a sequence of pieces of information from a sender, while hiding the selection of information from the sender and hiding the rest of the information from the receiver. In  1-out-of-2 OT, denoted $\myOT{2}{1}$, the sender has two strings $s_0,s_1$ and transfers $s_b$ to the receiver, where 
the receiver selects $b\in \{ 0,1 \}$ and the following two conditions hold: 
(1)~the sender does not know the value of $b$, and 
(2)~the receiver does not learn anything about $s_{1-b}$.

We use the generalization 1-out-of-$n$ OT, denoted $\myOT{n}{1}$: the sender has $n$ strings and transfers one string to the receiver, without knowing which string it transferred, and the receiver does not learn anything about the other $n-1$ strings.
An {ideal implementation} of OT might use a trusted third party: after obtaining the strings from the sender, and the index choice from the receiver, the party sends the chosen string to the receiver. 

OT can be implemented using public-key cryptography without a trusted third party. For example, 
$\myOT{2}{1}$ can be implemented as follows:
the receiver creates two random public keys but knows the private key corresponding to only one of them. The receiver sends the two public keys to the sender. The sender encrypts each string with a different public key and sends the resulting ciphertexts to the receiver. 
Because the receiver knows the private key corresponding to only one of the public keys, 
the receiver can decipher only one of the strings,
and the receiver will learn nothing about the other string.

Implementing OT with public-key operations, however, is computationally expensive. 
Seeking faster implementations, researchers have explored the possibility of implementing OT using symmetric-key cryptography, but Impagliazzo and Rudich~\cite{impagliazzo1989limits} showed that it is unlikely to find black-box constructions of OT using one-way functions. 
Beaver~\cite{beaver1996correlated}, however, shows that using one-way functions, a small number of \textit{base OTs} can be extended to any number of OTs. Namely, a small number of OTs that are built using expensive public-key operations can be extended to create a large number of OTs using symmetric cryptographic primitives. {After Beaver's seminal work, more efficient extensions of oblivious transfer have been introduced that are secure against passive adversaries~\cite{ishai2003extending,asharov2013more}, and secure against active adversaries~\cite{nielsen2012new,asharov2015more}.}

Seeking greater efficiency,
Bellare and Micali~\cite{Bellare89} created an $\myOT{2}{1}$ that requires two rounds. 
Naor and Pinkas~\cite{Naor01} reduced the number of exponentiations during run-time in Bellare and Micali from two to one on the sender's side. 
They also extended $\myOT{2}{1}$ to $\myOT{n}{1}$. 
In this $\myOT{n}{1}$ technique, the sender performs $n$ exponentiations in an initialization step, and uses the resulting values for all subsequent transfers.

Noar and Pinkas~\cite{naor2005compsec} showed how to extend any $\myOT{2}{1}$ protocol to an $\myOT{n}{1}$ protocol---with $O(n\log n)$ calls to $\myOT{2}{1}$---that provides sender
and receiver security computationally, if the underlying $\myOT{2}{1}$ provides sender and receiver security (see Section~\ref{sec:RACovert} for definitions). 
Among the most efficient OT protocols that are secure against active adversaries (including possibly
the sender and receiver)
are~\cite{peikert2008framework,chou2015simplest,asharov2017more}. 

Chou and Orlandi~\cite{chou2015simplest} computed more than $10,000$ $\myOT{2}{1}$s per second using one thread of an Intel Core i7-3537U processor.
Even with the overhead of building each $\myOT{n}{1}$ from $\myOT{2}{1}$s
using Noar and Pinkas's technique,
the $\myOT{n}{1}$s for message delivery in AOT should 
run sufficiently quickly for typical ACS applications.
Moreover, considering the horizontal scalability of Level-3 nodes, 
$\myOT{n}{1}$ should not be a bottleneck in AOT for message delivery for large numbers of users.

\mysavespace
\subsection{Applications of Oblivious Transfer} 

OT is a powerful primitive that alone can be used to implement any two-party or multiparty secure computation~\cite{kil88,crepeau1995committed}. 
We briefly point out some examples.

Nurmi {et al.}~\cite{nurmi91} used $\myOT{n}{1}$ to enable a trusted election authority
to distribute credentials to each of $n$ voters such that the election authority
does not learn the credential of any voter. 
Hence, when each voter uses their credential to cast their ballot,
the election authority cannot link ballots to voters. 

Even {et al.}~\cite{even85} used OT to sign contracts.
Two parties use $\myOT{2}{1}$ to exchange secrets, where
knowledge of the other's secret implies their commitment to the contract. 
Here, OT guarantees that
each party sends its secret correctly, and 
both parties simultaneously exchange their secrets.

Fagin {et al.}~\cite{fagin96} used OT to enable two parties to compare their secrets without
revealing them (e.g., a user wants to prove their identity using a password but does not trust the medium).

Javani and Sherman~\cite{javani2020bvot} use OT to
provide perfect ballot secrecy and ensure correct vote casting in a self-tallying boardroom voting protocol.

OT has also been used in mental poker~\cite{crepeau1986}, 
fair computation~\cite{goldwasser1990}, and zero-knowledge proofs~\cite{desantis1995}.


\mysavetinyspace
\section{Overview}

We explain AOT's communication model, adversarial model, goals, and
three-level architecture.

\mysavespace
\subsection{Communication Model} \label{sec:com-model}

Senders and receivers communicate with AOT using a trusted user application
(e.g., running on a smartphone or workstation). We refer to user applications simply as senders and receivers. 
We assume users have access to a public-key infrastructure and know 
the public keys of the other users and the AOT nodes.

Each sender sends a message and associated tag to AOT, which posts the tag on a public bulletin board.  Recipients detect tags related to them, and using
OT, download the associated messages.
{Each pair of users who want to communicate with each other must establish a shared secret}, which the sender uses to generate a tag. Users do not need to establish shared secrets with any of the mixnodes.

\mysavespace
\subsection{Adversarial Model} \label{sec:adversary}

We consider two types of adversaries: active and covert, each with the same
capabilities. Their common goal is to identify communicants, or to link senders and receivers of messages.  We do not consider denial-of-service attacks.

The adversary can be any of the users, AOT nodes, or a 
Dolev-Yao network intruder~\cite{Dolev15}.
The adversary cannot control all of the AOT users at the same time.

We assume authenticated and encrypted communications (e.g., TLS) between users and AOT, and among all nodes within AOT.
We assume that the adversary cannot defeat standard cryptographic functions.

An {\it active adversary} can monitor traffic between users and AOT, as well as 
traffic among nodes within AOT.  An active adversary can also 
delay incoming messages for an arbitrary amount of time, remove legitimate incoming messages, and inject arbitrarily many messages into the system.  
An active adversary can attempt to replay and modify messages.

A {\it covert adversary}~\cite{aumann2007security} seeks to keep the execution of any attack, and their involvement in it, undetected.

\mysavespace
\subsection{Goals}
Our design goal is an ACS that has the following properties:
\begin{enumerate}
    \item \textbf{Unlinkability.} The adversary should not be able to link senders and receivers of the communications. AOT provides receiver anonymity even if a covert adversary compromises the entire network (see Section~\ref{sec:anom-anl}).
    \item \textbf{Scalability.} The system should be able to scale with increasing numbers of users.
    \item \textbf{Integrity.} Either AOT delivers the messages unaltered to receivers, or it
detects message modification and identifies the source of the modification.
\end{enumerate}

\mysavespace
\subsection{Architecture} \label{sec:arch}

\begin{figure}[ht!] 
\centering
	\includegraphics[width=\columnwidth]{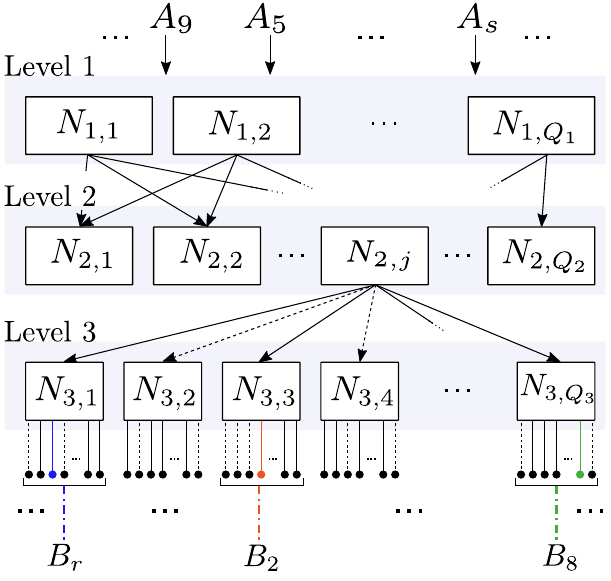}
	\caption{Summary of how AOT works.
	Senders send encrypted messages with associated tags
	to Level-1 nodes of AOT.  
	Level-1 nodes receive messages, remove the sender information from them,
	arrange the messages into batches of size $\beta_1$, shuffle the messages within each batch,
	and send the batches to Level-2 nodes.
	Level-2 nodes receive messages from Level-1 nodes, arrange them into batches of size $\beta_2$,
	add additional dummy messages to these batches, shuffle the real messages within each batch, 
	send the real messages to active nodes at Level-3, and send the dummy messages 
	(denoted by dashed lines) to passive nodes at Level-3.
	Level-3 nodes receive messages from Level-2 nodes, select collections of messages to publish, 
	and publish the tags of these selected messages 
	(denoted by small shaded circles and the end of lines). 
	Receivers detect tags related to them and request the associated encrypted payloads from Level-3 nodes using OT. Dummy messages help mitigate blending attacks, as explained in Section~\ref{sec:blending}.}
	\label{fig:overview}
	
\end{figure}

As explained in Figure~\ref{fig:overview}, 
AOT comprises a cascade of nodes organized in three levels. 
Each level performs a different function, and 
all nodes within the same level perform the same function.
Each level can have a different number of nodes.

Each level can scale ``horizontally'' (adding more nodes to that level) and independently of the other levels. Scaling nodes horizontally serves two purposes: (1)~to have higher availability and efficiency by distributing the load among multiple nodes, and (2)~to make it harder for the adversary to control the entire network.

Communications between levels are digitally signed and encrypted using public-key cryptography.


We denote each node as $N_{i,j}$, where $1\leq i \leq 3$ is the node's level, and 
$j$ is the node's position within that level.
For each Level $i$, let $Q_i$ denote the number of nodes
at Level $i$.

{\it Level-1 nodes.} Level-1 nodes receive messages, strip sender information from them, arrange messages into batches of size $\beta_1$, called {\it containers}, shuffle the messages within each container, and send the containers to Level-2 nodes.

{\it Level-2 nodes.} Each Level-2 node decrypts the messages that it receives from Level-1 nodes, arranges them in batches of size $\beta_2$, shuffles the messages within each batch, 
adds dummy messages, and sends the batches to Level-3 nodes. We assume $\beta_2=Q_1\beta_1$; however, $\beta_2 > \beta_1$ can have arbitrary values.
Dummy messages and dummy requests by users (see Section~\ref{sec:blending}) help mitigate blending attacks.



{\it Level-3 nodes.} Level-3 nodes enable receivers to retrieve messages sent to them.  For each round, Level-3 nodes organize themselves into ``active'' and ``passive'' nodes, and communicate this organization to Level-2 nodes.
Passive nodes receive the dummy messages; 
active nodes receive genuine messages. Each receives messages in
batches of size $\beta_2$ and delivers them.

\mysavetinyspace
\section{Anonymization by Oblivious Transfer} \label{sec:protocl}

We describe the AOT protocol (including 
how users send messages, how AOT processes them,
how AOT uses message tags,
and how it publishes messages),
the user app, and 
two-way communications using AOT.
Table~\ref{tab:notation} summarizes our notation.



\mysavespace
\subsection{Message Tags}  \label{sec:ud}

AOT requires each sender and receiver to have a {\it shared secret}. 
Users may establish these secrets anonymously using any method they
choose, including using AOT through a process called 
a {\it handshake} (see Section~\ref{sec:handshake}). 
Let $\sigma_{A,B}$ denote the shared secret between users $A$ and $B$. 
Communicants use the shared secrets to compute {\it tags}, which AOT posts
when making messages ready for delivery. Receivers recognize messages
by their tags; hence each tag should be identifiable
only by its intended recipient.

For each message sent from $A$ and $B$, users compute
a {tag} $k_{A,B} = f(\sigma_{A,B},c)$, where $c$ is a counter and
$f$ is a cryptographic key-derivation function (see \cite{krawczyk2010hkdf,HKDF}).
Specifically, $c$ denotes the number of successful communications between $A$ and $B$, where a successful communication is a communication in which the recipient of the message replies back to the sender, or sends an acknowledgment of receiving the message.
Section~\ref{sec:clnt} explains how AOT uses $c$, including 
how AOT handles synchronization issues.

\mysavespace
\subsection{Sending and Processing Messages}\label{sec:AOT}

\begin{figure*}[ht!] 
\centering
	\includegraphics[width=\textwidth]{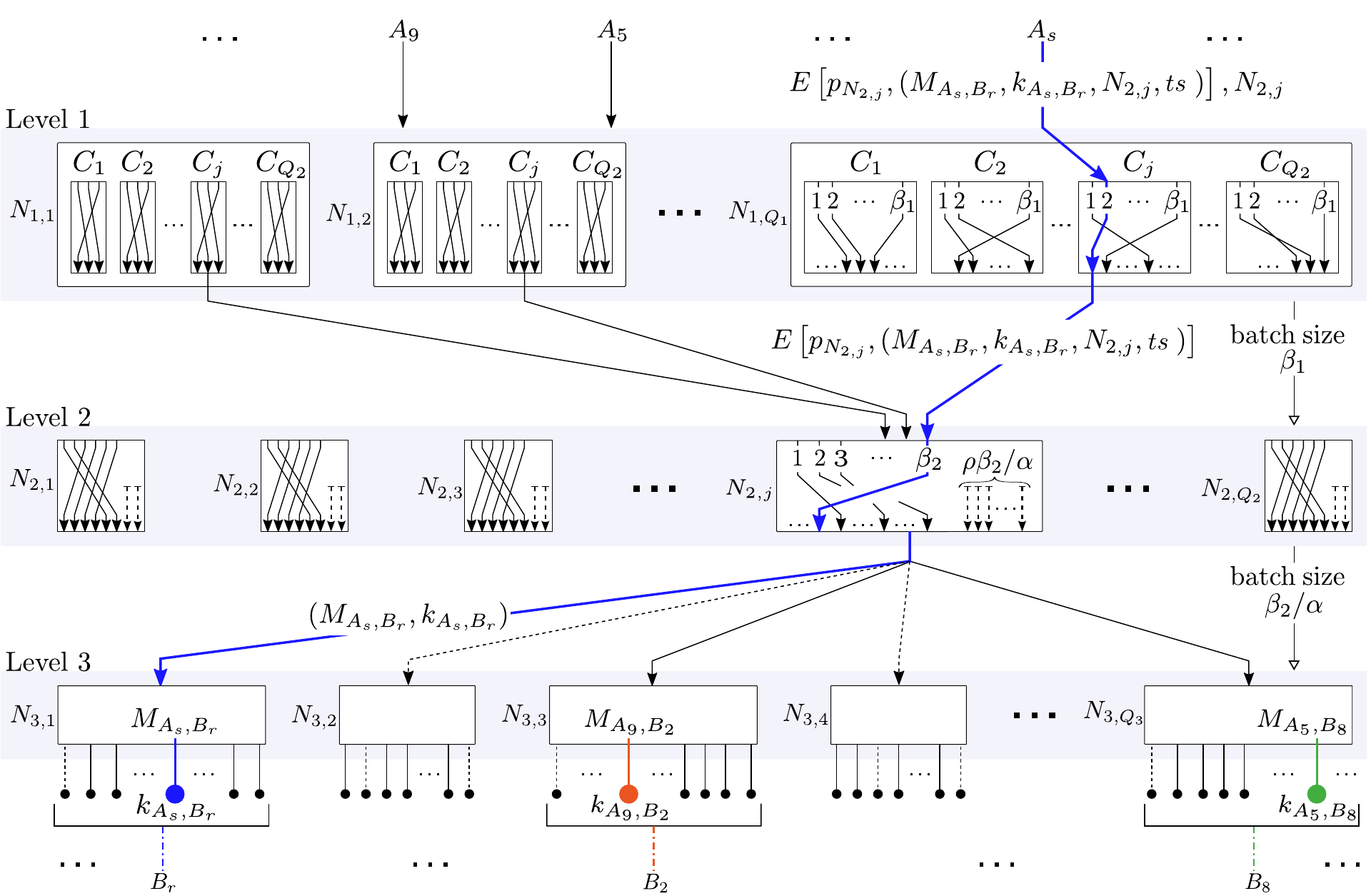}
	\caption{Detailed view of how messages flow through AOT.
	Each Level-1 node maintains a container, denoted $C_i$, for each Level-2 node~$N_{2,i}$.
	Crossing lines denote shuffling of messages.
	Level-2 nodes add dummy messages (denoted by dashed lines).
	Thick blue lines identify the route of a message from sender $A_s$ to receiver $B_r$. }
	\label{fig:flow}
\end{figure*}

Figure~\ref{fig:flow} shows how messages flow through AOT.
Suppose a sender $A$ wishes to send a payload $x_{A,B}$ to
a receiver $B$.  To begin, $A$ encrypts the payload by
computing

\begin{equation} \label{eq:M}
M_{A,B} = E\left[p_{B},(x_{A,B}, n) \right],
\end{equation}

\noindent where $p_B$ is the public key of $B$ and $n$ is a nonce.
$E[p_B,M]$ denotes {public-key encryption} of $M$ using the public key $p_B$---that provides confidentiality and ciphertext integrity---as described, for example, by Bernstein's~\cite{crypto_box} 
(see Appendix~\ref{Appendix-A}).\footnote{{Alternatively, 
instead of using public-key encryption, $A$ could encrypt the 
payload $(x_{A,B}, n)$ using symmetric-key encryption,
where the key is derived from the shared secret (in a fashion similar
to how AOT computes the message tags).
Nevertheless, $A$ would still have to encrypt 
their communication
using the public key of a Level-2 node. 
For consistency, we choose to use $E$ for both encryptions.}}

Next, $A$ chooses a Level-1 node $N_{1,i}$
and a Level-2 node $N_{2,j}$ at random, for some
$1 \leq i \leq Q_1$ and $1 \leq j \leq Q_2$.
Sender $A$ generates and sends the following message
to $N_{1,i}$:

\begin{equation}\label{eq:format}
m=  \left(
        E\left[
            p_{N_{2,j}},
            \left(M_{A,B},k_{A,B},N_{2,j},ts\right)
        \right],
        N_{2,j}
    \right) ,
\end{equation}

\noindent where $p_{N_{2,j}}$ is the public key of node $N_{2,j}$; 
the value $k_{A,B}$ is the tag of $M_{A,B}$;
and $ts$ is the timestamp for when $m$ was created. 
To prevent the adversary from linking incoming and outgoing messages based on their sizes, 
AOT uses a fixed size for all message payloads and tags.

For $1 \leq j \leq Q_2$, each Level-1 node maintains one {\it container} $C_j$ of messages for each Level-2 node $N_{2,j}$.
Upon receiving the messages from senders, each Level-1 node strips the sender information off the messages and puts them in the container corresponding to the specified Level-2 node.
Whenever any container is filled with
$\beta_1$ messages, the Level-1 node shuffles the order of messages in the container, sends the messages to the corresponding Level-2 node, and 
empties the container. 

For $1 \leq j \leq Q_2$,
node $N_{2,j}$ receives $\beta_1$ messages of the following form from a Level-1 node:

\begin{equation}\label{eq:N2}
        E\left[
            p_{N_{2,j}},
            \left(M_{A,B},k_{A,B},N_{2,j},ts\right)
        \right].
\end{equation}

\noindent
$N_{2,j}$ then decrypts these messages, puts them in {\it batches} of size $\beta_2$, and permutes the order of the messages within each batch.
Level-1 and Level-2 nodes commit to these permutations using a perfectly hiding commitment scheme~\cite{halevi1996}, broadcasting the commitments to the other nodes.

To prevent replay attacks, Equation~\ref{eq:format} includes a timestamp $ts$.
Each Level-2 node maintains a record of the (message, tag)-pairs of the messages that it has processed during the last $T$ minutes.
If they receive a message with an old timestamp that has been created within last $T$ minutes, they check the record of previously processed messages. If the message had been processed, they drop the message (see Section~\ref{std-attacks}).

Each \textit{round} is the interval during which Level-2 nodes receive $\beta_2$ messages and send them to the Level-3 nodes. For $l=1,2,\dots$, let $\mathcal{R}_l$ denote round $l$, and let $\Psi_l$ denote the batch processed at round $l$, where the first round is
$\mathcal{R}_1$.

During each round, AOT divides the Level-3 nodes into
$\alpha$ {\it active} nodes and $\rho$ {\it passive} nodes (see Section~\ref{sec:AP-division}).
Independently for each batch, the active nodes randomly partition the set of messages in the batch into $\alpha$ subsets.
Let $P$ denote the corresponding partition of message indices.
Each active node then receives one of the subsets of $\beta_2/\alpha$ 
messages from the partition.
AOT partitions the nodes into active and passive nodes so that a compromised Level-3 node cannot process all of the batches (see Section~\ref{sec:AP}).

The batch $\Psi_l$ that $N_{2,j}$ processes 
at round $\mathcal{R}_l$ comprises $\beta_2$ tuples of 
(encrypted payload, tag)-pairs:

\begin{equation}
\begin{split}
\Psi_l  =  \big\{ &  \left(M_{A_{s_1},B_{r_1}} ,k_{A_{s_1},B_{r_1}}\right), \\
          &         \left(M_{A_{s_2},B_{r_2}} ,k_{A_{s_2},B_{r_2}}\right), \\
          &         \qquad \vdots \\
          &         \left(M_{A_{s_{\beta_2}},B_{r_{\beta_2}}} ,k_{A_{s_{\beta_2}},B_{r_{\beta_2}}}\right) \big\} ,
\end{split}
\end{equation}

\noindent where $A_{s_1},B_{r_1}, A_{s_2},B_{r_2}, \dots, A_{s_{\beta_2}},B_{r_{\beta_2}}$ denote the senders and receivers of messages in the batch.

For each batch $\Psi_l$ that node $N_{2,j}$ processes,  $N_{2,j}$ creates $\rho\beta_2/\alpha$ {\it dummy messages}. 
As specified by partition $P$, 
node $N_{2,j}$ sends $\beta_2/\alpha$ real messages from the batch to each active Level-3 node, and $N_{2,j}$ 
sends $\beta_2/\alpha$ dummy messages to each passive node. 
We call each of these sets of $\beta_2/\alpha$ messages, real or dummy, 
{\it buckets} and denote each bucket by $\Phi$.
Although dummy messages increase the communication load among the mixnodes, they do not create any additional computational load because users do not ask to receive dummy messages,
except possibly in some dummy requests.

\mysavespace
\subsection{Publishing Messages}
\label{sec:publish}

AOT enables receivers to retrieve messages through two steps:
{\it publication} and {\it delivery}.
Each Level-3 node maintains a {\it message repository},
{\it publication repository}, and public
{\it bulletin board}.  During publication, the Level-3 node
moves messages from its message repository into its publication
repository and posts the associated message tags on its bulletin board.
During delivery, the recipient engages the Level-3 node
in an OT to retrieve the messages it recognized from
their associated tags.

\begin{figure*}[h] 

\begin{subfigure}[t]{.47\textwidth}\centering

\begin{framed}

\begin{algorithmic}[1]

\Procedure{$\mathcal{P}$}{$\tau $, $\lambda$, $j$} \Comment{Collect messages to publish}
\State  $M \leftarrow $ Empty Message Repository 
\State  Add dummy messages to $M$
\While{Node is alive}

\State Add the new bucket $\Phi_{\lambda , j}$ to $M$

\Loop \: every $\tau/\lambda$ seconds 

\State $L \leftarrow $ Empty set of messages to be published

\For{$1\leq i \leq \lambda$ }
\State Choose $\beta_2/(\alpha \lambda)$ messages at random from bucket $\Phi_{i,j}$ in $M $
\State Add chosen messages to $L$ 
\State Remove chosen messages from  $\Phi_{i,j}$
\EndFor
\State \textbf{Publish} $L$
\EndLoop

\EndWhile

\EndProcedure
\end{algorithmic}
\end{framed}

\caption{Procedure $\mathcal{P}$ used by each Level-3 node to collect a list of
messages to publish.  Each node $N_{3,j}$ randomly selects messages from each of its
past $\lambda$ buckets $\Phi_{1, j} , \ldots , \Phi_{\lambda , j}$.
The procedure guarantees that every incoming message 
is published in less than $\tau$ seconds from its arrival.}
\label{alg:msgrepo}
\end{subfigure}%
\hfill
\begin{subfigure}[t]{.47\textwidth}\centering
\begin{framed}
  \includegraphics[width=\linewidth]{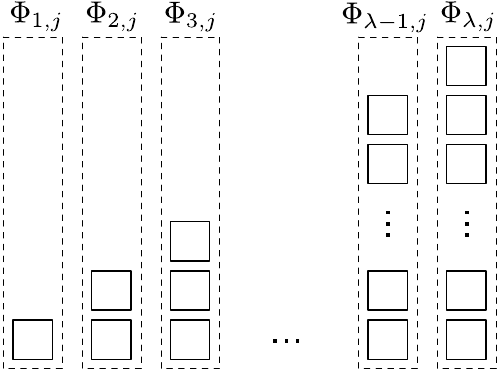}
  \end{framed}

  \caption{Message repository for each Level-3 node.  
  Representative node $N_{3,j}$ maintains
  a message repository of the $\lambda$ most recent buckets 
  $\Phi_{1, j} , \ldots , \Phi_{\lambda , j}$ of messages received,
  where $\Phi_{\lambda , j}$ is the most recent bucket. 
  Each square denotes $\beta_2/(\alpha \lambda)$ messages. 
  For each $1\leq i \leq \lambda$, bucket $\Phi_{i,j}$ (denoted by dashed lines)
  has $i\beta_2/(\alpha \lambda)$ messages; the repository has a total of
  $\beta_2(\lambda+1)/(2\alpha)$ messages.}
  \label{fig:repo}
\end{subfigure}

\caption{How each Level-3 node collects messages from its message repository
to publish in the AOT bulletin board. To increase the anonymity set and to obfuscate 
the flow of the messages within the system, each Level-3 node collects 
messages from its $\lambda$ most recent buckets of messages received.}
\label{fig:test}
\end{figure*}

As explained in Figure~\ref{fig:repo}, upon receiving buckets
of messages from Level-2 nodes, each Level-3 node puts
messages into its message repository.  
The repository consists of messages from the latest $\lambda$ buckets received.

During the {publication step}, each Level-3 node chooses $\beta_2/\lambda\alpha$ messages randomly from each of the $\lambda$ buckets in its message repository and moves them to its {publication repository}. The node also posts their tags on its bulletin board
in a {\it publication list}. Thus, Level-3 nodes do not deliver messages in exactly the same order in which they receive them from Level-2 nodes. 

Each Level-3 node maintains $\gamma$ messages in its publication repository, deleting the oldest messages as needed.
Recipients do not need to be online when their messages reach the
publication repository; they can retrieve their messages when they
return online (provided the messages still remain).

Figure~\ref{alg:msgrepo} shows how a Level-3 node moves messages from
its message repository into its publication repository.
The node stores each message in its message repository for less than $\tau$ seconds. 
If the rate of incoming messages to AOT is $\beta_2\tau/\lambda$ messages per second, with $\lambda$ batches in each message repository, 
and if Level-3 nodes perform their publication step every $\tau/\lambda$ seconds, the maximum delay caused by the publication step
is $\tau$ seconds for each message. 
The parameters can be adjusted for any rate of incoming messages.


During the delivery step, a receiver recognizes a tag in the publication list of a Level-3 node. Using OT, the receiver asks for the message corresponding to the tag from the Level-3 node that published the tag. The receiver engages in an $\textrm{OT}^{\zeta}_1$ 
session with the node and receives the message corresponding to the 
chosen tag. Here, $\zeta \leq \gamma$ is a parameter set by AOT (see Section~\ref{sec:gamma}).

When delivering the chosen message, the Level-3 node
sends the encrypted payload and tag together with a signed 
{\it message authentication code (MAC)} of the 
encrypted payload and tag. This MAC is
necessary to assure message integrity (see Section \ref{sec:integrity}).

In case of node failures, messages could be processed by other nodes at the same level. If a node at Level~1 fails, senders resend the messages to another Level-1 node. If an active Level-3 node fails, the Level-2 node that processed the batch sends the messages to another Level-3 node. 
If a Level-2 node fails, the sender recreates the message for another Level-2 node;
the need to resend the message in this case also arises in similar situations for
fixed cascades and free-routing mix networks.

\mysavespace
\subsection{User App and Its Roles}
\label{sec:clnt}

Senders and receivers interact with AOT through a trusted user app
running on the user's machine (possibly a smartphone).
This app helps the user carry out the AOT protocol, manage 
user keys and counters, and perform security-enhancing tasks.

The app facilitates message delivery by monitoring 
the bulletin boards.  When it notices a recognizable
tag, it initiates an $\textrm{OT}^{\gamma}_1$ with the
associated Level-3 node.

For each communicant, the app manages the associated 
shared secret and message counter, which it uses to compute tags.
The message counters of senders and receivers might lose
synchronization due, for example, to a dropped message
or failed delivery. To deal with such possible issues, the app can
try all values of the counter within some range (e.g., 
current value plus or minus up to some constant $\xi$,
say $\xi = 2$).

The app helps senders perform two integrity checks.
First, after sending each message, the sender checks that
the associated tag has been posted on some bulletin board.
The sender informs the appropriate Level-1 node of the
result of this check. Because the user's app monitors the bulletin board for coming messages, 
this check does not add any additional overhead. {As a result of this check, each message in AOT is a loop message~\cite{piotrowska2017loopix}, without requiring the sender to list their own address as the address of the receiver.}
Second, at random times, each sender asks for the messages that they have sent to the system to verify the system's integrity (see Section~\ref{sec:integrity}).
These times are
chosen at random from the interval~$[1 , T_1]$,  
The second check can also be performed by the app, 
using the same functionality for retrieving messages sent by others.
Users inform Level-1 nodes if any of these checks fail; see Section~\ref{sec:integrity}.

We assume that the rate of incoming messages to AOT is $\beta_2\tau/\lambda$ messages per second. AOT should receive $\beta_1Q_1Q_2/2$ on average before sending a batch of size $\beta_2$ to Level-3 nodes (see Section \ref{sec:anom-anl}). Therefore, senders should expect their messages to be published within $\tau+(Q_2\tau/(2\lambda))$ seconds.

The app also initiates dummy message requests to Level-3 nodes (see Section~\ref{sec:blending}). 
Each user makes a dummy request at random times, where the 
time to the next request is chosen at random with uniform distribution from the interval~$[1 , T_2]$.

\mysavespace
\subsection{Return Path, Delivery Acknowledgment, and Resending Messages}
\label{sec:return}

A strong property of AOT is that return paths can be implemented easily without any change in the communication model, enabling two-way anonymous communications. 
Because users ask for messages from AOT using OT, AOT does not require  the sender to specify the recipient's address. 
This property enables AOT to process each message in two-way communications in a stateless form, without requiring information about any previous message. 
None of the nodes in AOT can distinguish between forward and return messages. 
Upon receiving a message from a Level-3 node, the receiver can send
a return message as follows:
update the counter $c$ from previous communications with the sender, 
create a new tag, create the message tuple with new payload and tag, and send it to any Level-1 node.  Although AOT processes messages in a stateless fashion, senders and receivers operate in a stateful fashion
that requires knowledge of the counter value.

A recipient can acknowledge receipt of a message as follows.
They send a return message, as described above.  By observing the
resulting posted tag, the sender learns that the recipient received the
message, even if the sender does not retrieve the posted return message.

If a sender does not receive any acknowledgment of message receipt, 
the sender can resend the message.
Because recipients request messages using OT, Level-3 nodes do
not know which messages were received.
Known bounds on the latency caused by AOT,
and on the time tags remain in the bulletin board, 
facilitate such checking by senders.

\mysavespace
\subsection{Dividing Level-3 Nodes into Active and Passive Nodes}
\label{sec:AP-division}

Level-3 nodes partition themselves into $\alpha$ active nodes and $\rho$ passive node where $(1/3) \rho \leq \alpha \leq (3/4)\rho$ and $\alpha + \rho = Q_3$. 
Partitioning the Level-3 nodes has two phases: 1)~Initiation, and 2)~Division. 
Initiation happens when AOT first starts operating. 
Nodes perform the division phase for each round.

\mysavemorespace
\subsubsection{Initiation phase} 
\mysavelessspace
For each $1\leq j \leq Q_3$, node $N_{3,j}$ 
generates a random value $v_{3,j}$ and broadcasts a commitment of 
it to all \hbox{Level-2}
nodes and other Level-3 nodes. After each $N_{3,j}$ receives the commitment from all other nodes, it broadcasts $v_{3,j}$ to all Level-2 nodes and all other Level-3 nodes. Level-2 nodes and Level-3 nodes compute $\oplus^{Q_3}_{j=1} v_{3,j} $, where $\oplus$ denotes bit-wise XOR.

\mysavemorespace
\subsubsection{Division phase} 
\mysavelessspace
At each round $\mathcal{R}_l$, the Level-2 node that processes the batch, and all Level-3 nodes, compute\linebreak$V_l=g(\oplus^{Q_3}_{j=1} v_{3,j},l)$, where $g$ is a cryptographic key-derivation function (see \cite{krawczyk2010hkdf,HKDF}). 
Each node then divides $V_l$ into $Q_3$ bit strings of the same length, denoted by $V_{l,j}$, such that $V_l=V_{l,1}||V_{l,2}||\dots||V_{l,Q_3}$, where
$||$ denotes concatenation.
Each node creates the tuples $\left(V_{l,j},N_{3,j}\right)$ 
for all $1\leq j \leq Q_3$ 
and sorts the list of tuples based on their first elements. The first $\alpha$ nodes in the sorted list are actives nodes of the round.

The bit-length of $V_l$ should be set such that the probability of having $V_{l,j_1}=V_{l,j_2}$ where $j_1\neq j_2$ is negligible.

\mysavetinyspace
\section{Handshake Process} \label{sec:handshake}

We explain how the AOT protocol can be used by any two communicants to establish a shared secret anonymously, in what we call a {\it handshake process}. Any adversary that monitors these communications should be unable to identify the communicants or determine that
they engaged in a handshake process. 
This process can be used by senders and receivers, as needed, to establish the shared secrets needed for message delivery in AOT (see Section~\ref{sec:AOT}), as well as for any application unrelated to AOT.  The handshake process leverages the fact that the communicants
know each other's public keys.

For a sender $A$ to establish a shared secret with a receiver $B$, the
sender generates the message 

\begin{equation}\label{eq:shello}
\left(
        E\left[
            p_{N_{2,j}},
            \left(
                E\left[
                    p_{B}, ( B, R_A,ts)
                \right],
                k_{HS},N_{2,j},ts
            \right)
        \right],
        N_{2,j}
    \right) ,    
\end{equation}

\noindent
and sends it to AOT as they would any other message.  All such messages
include the same global constant tag $k_{HS}$, generated by AOT and 
sent to all users.
Here, $B$ is the receiver's identity and $R_A$ is a random value generated by $A$. 

Whenever any user application detects the special tag $k_{HS}$ 
on a bulletin board, 
they ask for the corresponding message 
$E\left[p_{B}, (B, R_A,ts) \right]$ using OT, 
but only $B$ can decrypt the message. The receiver $B$ decrypts the message and computes $\sigma_{A,B}=h(R_A)$ as the shared secret of $A$ and~$B$,
where $f$ is a cryptographic hash function.

Henceforth, $A$ and $B$ can communicate through AOT. 
For example, using the session shared key 
$k_{B,A} = f(\sigma_{A,B},1)$, $B$ can reply to $A$ 
with the message

\begin{equation}\label{eq:reply}
\left(
        E\left[
            p_{N_{2,j}},
            \left(
                E\left[p_{A},(x_{B,A},n)
                \right],
                k_{B,A}, N_{2,j}, ts
            \right)
        \right],
        N_{2,j}
\right) .
\end{equation}

\noindent If a sender and a receiver already share a suitable value, they do not need to perform the handshake.

AOT treats messages with the special tag $k_{HS}$ the same as all other messages. 
Therefore, any OT session for the handshake process 
is indistinguishable from any other OT session, except
that anyone can recognize how many special tags
were posted on the bulletin board.




\mysavetinyspace
\section{Security Notes} \label{sec:sec-anl}

We explain how AOT resists blending attacks and other standard attacks against mixnets.  We also explain how AOT achieves message authenticity and protocol integrity.  Section~\ref{sec:anom-anl} defines the security of OT and 
analyzes its anonymity, including stating and proving
its receiver anonymity property.

\mysavespace
\subsection{Resisting Blending Attacks}
\label{sec:blending}

AOT resists blending attacks with
dummy messages generated by Level-2 nodes and dummy requests initiated by users. In a {\it blending} (or $n-1$) {\it attack}~\cite{serjantov2002},
an active adversary attempts to determine the receiver of a message 
by allowing only the targeted message into the system---the adversary blocks or delays all other messages, or fills the system with its own messages. The adversary then tries to  determine the receiver by observing
who receives the targeted message.

Let
$m_{XY}=\left(
        E\left[
            p_{N_{2,j}},
            \left(M_{X,Y},k_{X,Y}, N_{2,j},ts\right)
        \right],
        N_{2,j}
\right)$
be the targeted message. The adversary knows that user $X$ generated and sent message $m_{XY}$; the adversary 
does not know the ciphertext $M_{X,Y}$ or the tag $k_{X,Y}$.

Without dummy requests, the adversary could identify the receiver as the
sole user to retrieve a message from the current  publication lists.
Without dummy messages, the adversary could identify---by the process of elimination---the sole message referenced in these lists not created by the adversary.

With dummy messages, the adversary could infer only that the targeted message is one among several (the targeted message and the dummy messages).
Furthermore, with dummy requests, the adversary could infer only that
the receiver is one among many (the receiver and any user who made a dummy request).\footnote{The chance that the tag of a real message collides with the tag of a dummy message is negligible.}

Thus, under a blending attack, the adversary can reduce the size of
the receiver anonymity set size for the targeted message to 
the number of dummy requests initiated by users 
during the time that the targeted message is in a publication list.

For example, if $U$ users make dummy requests every $T_2$ minutes, with $Q_3$ level three nodes that hold each message for $H$ hours in their publication repositories, AOT will receive $HU/(T_2Q_3)$ dummy requests during the time that any message is in a publication repository. Considering that each user will make more than one dummy request in $H$ hours, the adversary cannot reduce the receiver anonymity set.

\mysavespace
\subsection{Resisting Standard Attacks}\label{std-attacks}

We explain how AOT resists replay, traffic-analysis, tagging, and
intersection attacks.

\mysavemorespace
\subsubsection{Replay attacks} 
\mysavelessspace
AOT resists replay attacks because Level-2 nodes detect and drop any recent message that has been previously processed. Level-2 nodes drop the messages that are not created within the last $T$ minutes. Level-2 nodes also drop any message that has been created within the last $T$ minutes and has a (payload, tag)-pair that is equal to a (payload, tag)-pair of some previously processed message. If a user wants to resend their message, {they recreate the message with the same encrypted payload, a new tag (with a new counter---see Section~\ref{sec:clnt}), and a new timestamp.}

\mysavemorespace
\subsubsection{Traffic-analysis attacks} 
\mysavelessspace
AOT resists traffic-analysis attacks through standard
message and batch sizes and message reordering.
AOT accepts only messages of a fixed size.
Each Level-1 node sends messages in batches of size $\beta_1$,
and each Level-2 node sends batches of size $\beta_2/\alpha$ (dummy or real).
Furthermore, each Level-1 and Level-2 node reorders all messages
within each batch.

\mysavemorespace
\subsubsection{Tagging attacks} 
\mysavelessspace
In a tagging attack, the adversary modifies (``tags'') a message before it enters the anonymity network and observes the output messages to recognize the modified message.  The only published outputs of AOT are the message tags, which senders compute from shared secrets that the adversary does not know.  If the adversary could replace a message with a modified one that had a different tag, the intended recipient would not recognize the modified tag, and the adversary would gain no useful information.
Moreover, since messages are encrypted using {an encryption} algorithm that provides ciphertext integrity, the adversary cannot alter the messages without being detected.

\mysavemorespace
\subsubsection{Intersection attacks} 
\mysavelessspace
AOT resists intersection attacks through dummy requests.
In an intersection attack~\cite{danezis2004statistical}, the adversary attempts to break the anonymity of communications by 
correlating the behaviors of users.  For example, 
if a receiver $B$ always receives a message whenever the sender $A$ sends one, the adversary might conclude that $A$ and $B$ are communicating. 
Mitigating intersection attacks is challenging and many anonymity systems
are vulnerable to such 
attacks~\cite{kedogan2002limits,chaum2017cmix,wolinsky2012dissent,kwon2016}. 
In AOT, dummy requests (including by actual receivers) help hide the correlations of user
behaviors from the adversary.
As explained in Sections~\ref{sec:blending} and~\ref{sec:RACovert}, the recipient of any published message is indistinguishable from the users who send dummy requests.

\mysavespace
\subsection{Protocol Integrity} \label{sec:integrity}

We explain how entities detect certain possible integrity errors and what they do about such errors.  Such error detection is possible because users encrypt their payloads using {an encryption algorithm that provides ciphertext integrity}, and 
nodes protect their communications with other nodes using TLS.
As in cMix~\cite{chaum2016cmix}, we define the protocol integrity as follows: 

\begin{definition}\label{def:integrity}

A protocol maintains integrity if at the end of each run involving honest
users:
\begin{enumerate}
    \item either, all the messages are delivered unaltered to the intended recipients;
    \item or, a malicious mixnode is detected with a non-negligible probability, and no honest party is proven malicious.
\end{enumerate}

\end{definition}

If the users and the mixnodes in AOT follow the protocol correctly, 
AOT delivers messages anonymously to recipients without modification. 

Whenever users or mixnodes detect an altered message, AOT will begin an \textit{audit process}. During the audit process AOT traces the message from the step during which the alteration is detected, back to the sender of the message. So tracing a message is possible because the Level-1 and Level-2 nodes commit to the shuffles that they apply to messages in containers and batches. 
Because AOT delivers messages using OT, tracing an altered message to its sender does not
violate anonymity. The adversary cannot detect the receiver of a message, nor can it link the sender and the receiver of a message by incorrectly claiming that the message has been altered.

We consider three scenarios where users and/or AOT nodes act maliciously.

\textbf{Level-2 nodes detect an altered message.} 
Each \hbox{Level-2} node checks each in-coming user message for integrity. 
If the integrity check fails, the Level-2 node asks the
forwarding Level-1 node to prove that it has not modified the message.
If the Level-1 node can prove
that it did not modify the message, then AOT assumes that the 
input received by the Level-1 node was in error (e.g., by the sender
or in transit). 

\textbf{AOT publishes an altered message.} 
Each receiver should check the received message for integrity.
If a receiver detects an integrity error
caused by a malicious node, the receiver can notify the sender using AOT.
The sender can ask for the message and notify the Level-1 node that received 
the message. When users notify AOT that an altered message has been published, nodes will then trace the message and detect the malicious node.

Moreover, after a sender sends a message, they can ask for the message and verify if it has been altered. Senders also send dummy messages to verify the integrity of the system.

Level-3 nodes send a signed hash of the encrypted (payload, tag)-pair, along with the tuple in the delivery step. Therefore, Level-3 nodes cannot alter the message during the publication and delivery steps, or use the unaltered message during an audit, without being detected.  

\textbf{Honest parties cannot be proven malicious.} 
Senders encrypt messages with {a public-key encryption 
system that provides ciphertext integrity.} 
Therefore, malicious nodes cannot alter a message and incorrectly claim that 
some user modified the message.
Moreover, all of the communications between mixnodes and users, and among the mixnodes, are signed. If any party is accused of altering a message, they can use the signed messages that they receive from other parties as proof of their correct behavior. 


\mysavetinyspace
\section{Anonymity Analysis} \label{sec:anom-anl}

We analyze the anonymity of AOT in two ways.
First, we bound the size of the anonymity set.
Second, we show that AOT has receiver anonymity even if
all nodes are compromised.

\mysavespace
\subsection{Receiver and Sender Anonymity Sets}

We bound the size of the sender anonymity set and receiver anonymity set in AOT. We do so without considering that the adversary may be able to reduce the size of the anonymity sets---though we are not aware of any such attacks. For example, the adversary may be able to reduce the size of the anonymity sets if they have additional information about the communication (e.g., time or location), or if the protocol has vulnerabilities. As such, our calculations should be interpreted as upper bounds on the sizes of the anonymity sets.
With similar caveats, we also show that members within the anonymity sets are equally like to be the sender or the receiver.

In Section~\ref{sec:pool}, we also analyze the anonymity of a version of AOT without the latency limit $\tau$ using an information-theoretic definition of anonymity.


Chaum~\cite{chaum1988dining} introduced the concept of anonymity set
in 1988.
As defined by Pfitzmann and Kohntopp~\cite{pfitzmann2001}, {\it anonymity}
``is the state of being not identifiable within a set of subjects, the anonymity set.'' 
Kesdogan {et al.}~\cite{kesdogan1998} define
anonymity set as the set of users that have a non-zero probability of being the sender or receiver of a particular message. 
Pfitzmann and Kohntopp argue that stronger anonymity is associated with 
a larger anonymity set and an even chance of being the sender or receiver
among members of the set.

We assume that messages arrive into AOT at the rate of $\beta_2\tau/\lambda$ messages per second, and that on average, Level-2 nodes receive an equal number of messages per second from Level-1 nodes. We also assume that, when Level-2 nodes first process and send real messages to
Level-3 nodes, each of the Level-3 nodes has dummy messages in their repository.

As explained in Figure~\ref{alg:msgrepo}, during
the publication step, each Level-3 node publishes $\beta_2/\alpha$ tags of messages that they randomly select from the $\lambda$ most recent buckets---$\beta_2/(\alpha \lambda)$ from each bucket---for publication from the node's message repository. 
Altogether, Level-3 nodes publish $\beta_2+ \rho\beta_2/\alpha$ messages. Assuming receivers select active and passive nodes uniformly at random for each round, on average $\rho\beta_2/\alpha$ of the selected messages are dummy messages.


\mysavemorespace
\subsubsection{Sender Anonymity Set} 
\mysavelessspace
Each published tag could have been selected from $\lambda$ recent 
batches of size $\beta_2$ that Level-2 nodes processed. 
We explain:
for $1\leq i \leq\lambda$, let $\Phi_{i,j}$ denote the $\lambda$ buckets in a Level-3 node $N_{3,j}$, where $\Phi_{\lambda, j}$ denotes the most recent bucket. 
As explained in Figure~\ref{fig:repo},
before any publication, the message repository of active node $N_{3,j}$ contains $i\beta_2/(\alpha \lambda)$ messages from buckets $\Phi_{i,j}$, for all $1\leq i \leq\lambda$.
During each publication step,
node $N_{3,j}$ chooses $\beta_2/(\alpha \lambda)$ messages from each $\Phi_{i,j}$.
Therefore, $N_{3,j}$ selects
messages from $\lambda$ recent batches received from Level-2 nodes.  


Level-1 nodes accumulate incoming messages in the containers $C_j$ before sending them to the Level-2 nodes $N_{2,j}$, for $1 \leq j\leq Q_2$. 
Assuming that senders choose Level-1 nodes and Level-2 nodes uniformly at random, an incoming message can go to any of the $Q_1Q_2$ containers with approximately equal probability of $\approx1/(Q_1Q_2)$.
Containers will accumulate messages at a similar rate. 
Each container should have $\beta_1$ messages 
before its messages are sent to a Level-2 node.

Therefore, on average, Level-1 nodes will receive $\beta_1Q_1Q_2/2$ incoming messages before sending $\beta_2$ messages to a Level-2 node
in batches of size $\beta_1$.
Whenever any \hbox{Level-3} node publishes a message, it came from
$\lambda$ recent Level-2 batches.  Therefore, 
the size of the sender anonymity set in AOT is $\lambda\beta_1Q_1Q_2/2$. 

\mysavemorespace
\subsubsection{Receiver Anonymity Set} 
\mysavelessspace
Because OT hides the messages that receivers request, for any Level-3 node and for any message in its publication list, 
any receiver that initiates an OT session with the node
while the message is in the list
could be the receiver of the message. 
Assuming that receivers ask for all messages that AOT publishes, 
for each Level-3 node, the receiver anonymity set is at most $\gamma$
(the maximum number of messages in the publication lists) 
plus the number of dummy requests to the node during the same time period (minus the number of users who make real requests to receive messages as well as dummy requests).
In Section~\ref{sec:RACovert}, we show that the adversary cannot reduce the receiver anonymity set of any message below the number of dummy requests made to the publishing node.

\mysavespace
\subsection{Receiver Anonymity with Corrupted Nodes} 
\label{sec:RACovert}

Adapting definitions from Naor and Pinkas~\cite{naor2005compsec}, we
state definitions of receiver and sender security for OT.

\begin{definition}\label{def:rec-sec}
\textit{Receiver security in OT.} An $\textrm{OT}^m_1$ provides receiver security if and only if, for any probabilistic\linebreak
polynomial-time sender $\mathcal{A}$ with $m$ 
strings $s_1, s_2, \dots s_m$,\linebreak
given any $1 \leq i < j \leq m$ where the receiver chose either $s_i$ or $s_j$,
$\mathcal{A}$ cannot distinguish whether the receiver chooses $s_i$ or $s_j$.
\end{definition}

Sender security is defined in terms of a comparison between 
the information the receiver learns in the ideal implementation of OT
and the information the receiver learns in the real implementation.

\begin{definition}\label{def:sen-sec}
\textit{Sender security in OT.} 
An $\textrm{OT}^{\lambda m}_1$ provides sender security if and only if,
for every probabilistic\linebreak
polynomial-time receiver $\mathcal{B}$, 
substituting $\mathcal{B}$ in the real implementation of the protocol, 
there exists a probabilistic\linebreak
polynomial-time machine $\mathcal{B^{\prime}}$  
for the receiver’s role in the ideal implementation
such that, for every sequence of strings $s_1, s_2, \dots s_{\lambda m}$ of the sender, 
the outputs of $\mathcal{B}$ and
$\mathcal{B^{\prime}}$ are computationally indistinguishable.
\end{definition}

OT protocols introduced in~\cite{peikert2008framework,chou2015simplest,asharov2017more} satisfy Definitions~2 and 3 against malicious senders and receivers.

A strength of AOT's design is that it preserves receiver anonymity, even
if a covert adversary corrupts all of the nodes. This property follows from
three sources:
receivers retrieve messages using OT;
receivers issue dummy message requests; and
Level-1 nodes verify that their messages have been posted.

For example, a malicious Level-3 node might attempt to find the recipient of a targeted message by posting only the targeted message, along with a set of $\gamma-1$ fake messages. This attack would fail for two reasons:
First, some receivers would issue dummy requests.  Therefore, the Level-3 node could not uniquely identify the receiver. 
Second, because senders always verify that their messages have been posted,
senders would notice that their messages were not posted.

These observations lead to the following theorem.

\begin{theorem}\label{thm:rac}
If the oblivious transfer protocol used in AOT provides receiver security according to Definition \ref{def:rec-sec}, AOT provides receiver anonymity as defined in 
Definition~\ref{def:rec-anon} (below), 
even if a covert adversary controls all the nodes in AOT.
\end{theorem}



Section~\ref{sec:proof} presents a proof of Theorem~1 using
a game between a challenger and an adversary.



\mysavetinyspace
\section{Comparison with Other Anonymity Systems} \label{sec:compare}

We compare design features and security and efficiency properties of AOT with those of selected other ACSs: original mixnet (OM)~\cite{chaum81}, cMix~\cite{chaum2017cmix}, Riffle~\cite{kwon2016}, Stadium \cite{tyagi2017stadium}, Vuvuzela~\cite{van2015vuvuzela}, and Loopix~\cite{piotrowska2017loopix}.
Table~\ref{tab:compare} summarizes the comparison.

AOT has two novel design features, which contribute to its
security and performance.
First, each level in the cascade of nodes performs a different function.
Second, during each round, AOT assigns each Level-3 node an active or passive role, where AOT sends batches of dummy messages to passive nodes.

Systems that resist active attacks use dummy messages and dummy requests in a variety of ways.
To hide communication patterns, 
many systems that resist active attacks require users to engage in every round of communication, even if the users do not send or receive genuine messages
in every round.
In AOT and Loopix, however, users do not have to participate in every round. 

Most of the systems in our comparison group are scalable to support a large number of users (e.g., Loopix). Riffle is designed for communications of users within groups.
AOT (see Section~\ref{sec:horizontal}), Stadium, and Loopix scale horizontally.

 
\begin{table*}[ht!]

\centering
\caption{Comparison of AOT with selected ACSs with respect to
design characteristics, security, and efficiency.
The comparison set comprises
original mixnet (OM)~\cite{chaum81}, cMix~\cite{chaum2017cmix}, 
Riffle~\cite{kwon2016}, Stadium \cite{tyagi2017stadium}, 
Vuvuzela~\cite{van2015vuvuzela}, and Loopix~\cite{piotrowska2017loopix}.}
\label{tab:compare}
\begin{subtable}{\textwidth}
\centering
\caption{Design characteristics and features of selected ACSs. Two unique properties of AOT are nodes with different functionality, and grouping of nodes for each batch into active and passive nodes.}
\label{tab:compare-char}
\begin{tabular}{rlccccccc}
& &\shortstack{Different \\ Functionality \\ per Level} & \shortstack{Includes\\Passive\\Nodes} & \shortstack{Dummy\\Requests}&  \shortstack{Dummy\\Messages}& \shortstack{Fixed\\Routes}& \shortstack{Horizontal\\Scalability}&\shortstack{Participation\\per Round}\\ \hline


OM          & 1981  & $\times$       &  $\times$    & $\times$      & $\times$      & $\checkmark$  & $\times$      & $\times$      \\\hline
Vuvuzela    & 2015  & $\times$       & $\times$     & $\checkmark$  & $\checkmark$  & $\checkmark$  & $\times$      & $\checkmark$  \\\hline
Riffle      & 2016  & $\times$       & $\times$     & $\checkmark$  & $\checkmark$  & $\times$      & $\times$      & $\checkmark$  \\\hline
cMix        & 2017  & $\times$       &  $\times$    & $\times$      & $\times$      & $\checkmark$  & $\times$      & $\times$      \\\hline
Stadium     & 2017  & $\times$       & $\times$     & $\checkmark$  & $\checkmark$  & $\times$      & $\checkmark$  & $\checkmark$    \\\hline
Loopix      & 2017  & $\times$       & $\times$     & $\checkmark$  & $\checkmark$  & $\times$      & $\checkmark$  & $\times$      \\\hline
AOT         & 2021  & $\checkmark$   & $\checkmark$ & $\checkmark$  & $\checkmark$  & $\times$      & $\checkmark$  & $\times$ \\
\end{tabular}

\end{subtable}

\vspace{8mm}

\begin{subtable}[h]{\textwidth}
\centering

\caption{Security and efficiency properties of selected ACSs. 
GPA denotes Global Passive Adversary, and 
RAC denotes Receiver Anonymity with Compromised network.
AOT provides receiver anonymity, even if a covert active adversary subverts all nodes.}
\label{tab:compare-prop}
\begin{tabular}{rlccccccc}
&&\shortstack{GPA \\ Resistance} & \shortstack{Active Attack \\ Resistance} & \shortstack{Scalable  \\ Deployment} &\shortstack{Low \\ Latency}  & \shortstack{Asynchronous \\ Messaging}  & \shortstack{Offline \\ Storage} & RAC \\ \hline
OM          & 1981  & $\checkmark $  &  $\times$     & $\times$     & $\times$       & $\checkmark$    & $\times$      & $\times$      \\\hline
Vuvuzela    & 2015  & $\checkmark $  & $\checkmark$  & $\checkmark$ & $\times$       & $\times$        & $\times$      & $\times$  \\\hline
Riffle      & 2016  & $\checkmark $  & $\checkmark$  & $\times$     & $\checkmark$   & $\times$        & $\times$      & $\times$  \\\hline
cMix        & 2017  & $\checkmark $  &  $\times$     & $\checkmark $& $\checkmark $  & $\checkmark$    & $\times$      & $\times$      \\\hline
Stadium     & 2017  & $\checkmark $  & $\checkmark$  & $\checkmark$ & $\times$       & $\times$        & $\times$      & $\times$    \\\hline
Loopix      & 2017  & $\checkmark $  & $\checkmark$  & $\checkmark$ & $\checkmark$   & $\checkmark$    & $\checkmark$  & $\times$      \\\hline
AOT         & 2021  & $\checkmark $  & $\checkmark$  & $\checkmark$ & $\checkmark$   & $\checkmark$    & $\checkmark$  & $\checkmark$ \\
\end{tabular}

\end{subtable}

\vspace{4mm}

\end{table*}

All of the systems in our comparison group resist {\it global passive adversaries (GPA)}, who observe all traffic between nodes of the system and between users and the system. We now describe how each system mitigates blending attacks.

Serjantov \textit{et al.}~\cite{serjantov2002} define two properties of {blending attacks}: an attack is \textit{exact} if the attacker can determine the receiver of a message with probability $1$; an attack is {\it certain} if the adversary can always isolate and trace the target message. 

cMix and OM follow the threshold mixing strategy; therefore, as noted by Serjantov \textit{et al.}~\cite{serjantov2002}, an adversary can perform an exact and certain blending attack on both systems.
Through sending loop messages, Loopix renders blending attacks 
uncertain and inexact. With {loop messages} Loopix determines if the adversary is blocking incoming traffic, enabling its nodes to change their behavior when under such attack.
Since loop messages are indistinguishable from normal messages, 
the most an adversary can do is to guess which messages are loop messages.

Vuvuzela and Stadium resist blending attacks by having all users make dummy requests in every round. 
Because sending messages is indistinguishable from receiving messages, to perform a blending attack in these two systems, the adversary's attempt to block incoming traffic also blocks the targeted message from reaching its intended recipient. 
In both systems, however, the adversary is able to reduce the anonymity set to a smaller group of communicants as follows.
To begin, the adversary hypothesizes the composition of some subgroup
of users who are communicating frequently.
The adversary blocks all incoming messages, except for those that originate from within this subgroup.  If the majority of the message locations in the memory of the
last mixnode in the cascade are accessed twice---once for writing and once for reading a message---the adversary can conclude that users within the subgroup are frequently communicating with each other.


AOT and Loopix are the only systems that provide offline message delivery
and that do not rely on synchronous rounds for security.
Synchronous rounds require mandatory user participation in every round.

As shown in Section~\ref{sec:RACovert}, 
AOT preserves receiver anonymity even if a covert adversary controls all nodes; none of the other systems has this property. 


\mysavetinyspace
\section{Discussion} \label{sec:discuss}

We explain our key design decisions, 
calculate the storage requirement for Level-3 nodes,
comment on the choice of the number of messages in each OT, 
and state some open problems.
Section~\ref{sec:pool} analyzes 
how converting AOT into a pool mix would affect anonymity and latency,



\mysavespace
\subsection{Key Design Decisions} \label{sec:decisions}

We discuss several important design decisions.

\mysavemorespace
\subsubsection{Using Oblivious Transfer} \label{sec:OTchoice}
\mysavelessspace
The central design decision was to use OT in message delivery.
At the cost of executing an OT protocol, OT contributes 
significantly to receiver anonymity: OT receiver security ensures that neither AOT nor the adversary learns which message the receiver retrieves.

An alternative design would be to post each message tag and its
encrypted payload $M_{A,B}$ on the bulletin board. In this alternative design,
the receiver could request messages with PIR~\cite{chor95pir} rather than with OT.  We chose to use OT rather than PIR because OT provides both
sender and receiver security, while PIR provides only receiver security.


{
Receiver security of OT increases the cost of the following two attacks:
\begin{itemize}
    \item[a)]  The adversary downloads a (message, tag)-pair,
re-encrypts it with the public key of a Level-2 node (adding a new timestamp),
and sends the ciphertext to the Level-2 node, while also blocking
new messages from reaching AOT with a blending attack. The adversary
observes which receivers request messages, in the hope of associating
a receiver with the replayed message.
\item[b)] The adversary monitors the frequency by which users initiate OT sessions with AOT. When a sender resends a message (when they do not receive an acknowledgment of delivery or a response to their message), the adversary notices a change in the number of OT sessions.
\end{itemize}
With sender security of OT, an adversary
must perform one OT for each (message, tag)-pair retrieved in this fashion.
With PIR, the adversary could download all (message, tag)-pairs at once, monitoring all published (message, tag)-pairs with, say, 
two OT sessions per day.}

\mysavemorespace
\subsubsection{Shared Secrets} \label{sec:shared}
\mysavelessspace
A key enabling design decision was to assume that each pair of communicants shared a secret, and to leverage this shared secret 
to enable each pair of communicants to recognize the tag of
messages sent between them. 
The cost is requiring each pair of communicants to establish
a shared key, and we show how AOT can be applied to do so. 

\mysavemorespace
\subsubsection{Multiple Nodes Per Level} \label{sec:horizontal}
\mysavelessspace
We designed AOT to scale horizontally for each level, permitting
multiple nodes per level. This decision enhances
load balancing, reliability, and security.  
The work of each level is spread over several nodes.
In case a node fails, other nodes on that level could process the messages.
To control a level, the adversary would have to control
several nodes on that level.

\mysavemorespace
\subsubsection{Message Batches at Level 1 and Level 2}
\mysavelessspace
Following a fundamental technique of mixnets, AOT processes
messages in batches, where AOT permutes the order of the
messages within each batch. Doing so enhances anonymity but can introduce delays.  Unlike most mixnets, AOT uses different
batch sizes at the first two levels; this property enables AOT to distribute the processing of messages between Level-1 and Level-2 nodes without introducing delays.

\mysavemorespace
\subsubsection{Number of Handshake Tags} \label{sec:HSchoice}
\mysavelessspace
Section~\ref{sec:handshake} proposes to use a single global handshake tag.  A consequence of this choice is that every user who initiates a handshake must request all messages with this tag.  
To reduce this overhead, one could use multiple handshake 
tags---for example, one per geographic region. 
This alternative choice would come at the cost of
revealing more information about communication patterns.

\mysavemorespace
\subsubsection{Active and Passive Nodes} \label{sec:AP}
\mysavelessspace
We chose, for each round, to partition Level-3 nodes into active and passive nodes, where active nodes handle real messages, 
and passive nodes handle dummy messages.  An alternative choice
would be for each Level-3 node to handle real messages and
to generate some additional dummy messages.
Our choice aims to obfuscate the internal flow of messages.


\mysavespace
\subsection{Storage Requirements for Level-3 Nodes} 
\label{sec:params}

We calculate the amount of storage required by each Level-3 node, for one choice of parameters.
Let $v$ denote the rate of incoming messages to the AOT (in Procedure $\mathcal{P}$, Figure \ref{alg:msgrepo}, we assume $v=\beta_2\tau/\lambda$ messages per second). 
AOT publishes $v + (\rho v/\alpha)$ real and dummy messages per second, where $\alpha$ is the number of active nodes
and $\rho$ is the number of passive nodes.
The rate of the incoming messages and published real messages are equal---regardless of number of previous batches $\lambda$ in the publication repository and the maximum delay $\tau$ for messages in the publication repository. 

For example, if AOT keeps the messages in each publication list for 12 hours before removing them, with $v=10,000$ messages per second, $300$ bytes message size, and $Q_3=5$, each Level-3 node will store
about $260$ gigabytes of data.



\mysavespace
\subsection{Number of Messages in Each Oblivious Transfer} \label{sec:gamma}

One of the parameter choices in AOT is the number of messages
$\zeta$ from which users select a message in each OT.
This choice affects the running time of OT and 
receiver anonymity.
Currently, we use $\zeta = \gamma$, the length of the publication list. 
An alternative decision would be to choose
$\zeta < \gamma$ and to allow each user to select from which
tags to choose in OT  
(e.g, set $\zeta = \gamma /10$).
This alternative decision
would permit a larger $\gamma$, when messages in the
publication list could be maintained for a longer time,
allowing users to be offline longer.  This choice, with its
shorter list of tags, might
also reduce the size of the receiver anonymity set.


\mysavespace
\subsection{Open Problems} \label{sec:open}

Open problems include:
(1)~Formally state and prove the security properties of AOT, and
(2)~perform a formal-methods analysis of the AOT protocol using
a protocol-analysis tool such as CPSA~\cite{SRP2020}.


\mysavetinyspace
\section{Conclusion}
\label{sec:conclude}

We introduced AOT, an anonymous communication system with mixnet architecture that provides offline message delivery and horizontal scalability. 
Through using oblivious transfer for message delivery,
AOT resists blending attacks, and 
provides receiver anonymity even if a covert adversary controls the entire network. 
The handshake protocol, which applies AOT to enable two communicants
to establish a shared key anonymously, is of independent interest.
AOT illustrates the power and flexibility of oblivious transfer 
as a building block in protocol design to enhance security properties.



\section*{Acknowledgments}
\label{sec:Acks}

We thank David Chaum and Jonathan Katz for helpful comments.
Sherman was supported in part by
the National Science Foundation under SFS grant DGE-1753681, 
and by the U.S. Department of Defense under 
CySP grants H98230-19-1-0308 and H98230-20-1-0384.


\bibliographystyle{acm}
\bibliography{main}

\begin{thebibliography}{10}

\bibitem{FJredacted}
REDACTED, unpublished document by one of the authors, 2021.

\bibitem{AdidaW07}
{\sc Adida, B., and Wikstr{\"{o}}m, D.}
\newblock Offline/online mixing.
\newblock In {\em {ICALP} 2007\/} (2007), pp.~484--495.

\bibitem{asharov2013more}
{\sc Asharov, G., Lindell, Y., Schneider, T., and Zohner, M.}
\newblock More efficient oblivious transfer and extensions for faster secure
  computation.
\newblock In {\em Proceedings of the 2013 ACM SIGSAC conference on Computer \&
  communications security\/} (2013), pp.~535--548.

\bibitem{asharov2015more}
{\sc Asharov, G., Lindell, Y., Schneider, T., and Zohner, M.}
\newblock More efficient oblivious transfer extensions with security for
  malicious adversaries.
\newblock In {\em Annual International Conference on the Theory and
  Applications of Cryptographic Techniques\/} (2015), Springer, pp.~673--701.

\bibitem{asharov2017more}
{\sc Asharov, G., Lindell, Y., Schneider, T., and Zohner, M.}
\newblock More efficient oblivious transfer extensions.
\newblock {\em Journal of Cryptology 30}, 3 (2017), 805--858.

\bibitem{aumann2007security}
{\sc Aumann, Y., and Lindell, Y.}
\newblock Security against covert adversaries: Efficient protocols for
  realistic adversaries.
\newblock In {\em Theory of Cryptography Conference\/} (2007), Springer,
  pp.~137--156.

\bibitem{beaver1996correlated}
{\sc Beaver, D.}
\newblock Correlated pseudorandomness and the complexity of private
  computations.
\newblock In {\em Proceedings of the twenty-eighth annual ACM symposium on
  Theory of computing\/} (1996), pp.~479--488.

\bibitem{Bellare89}
{\sc Bellare, M., and Micali, S.}
\newblock {Non-Interactive} oblivious transfer and applications.
\newblock In {\em Conference on the Theory and Application of Cryptology\/}
  (1989), pp.~547--557.

\bibitem{crypto_box}
{\sc Bernstein, D.~J.}
\newblock Cryptography in {NaCl}.

\bibitem{brickell2006efficient}
{\sc Brickell, J., and Shmatikov, V.}
\newblock Efficient anonymity-preserving data collection.
\newblock In {\em Proceedings of the 12th ACM SIGKDD International Conference
  on Knowledge Discovery and Data Mining\/} (2006), pp.~76--85.

\bibitem{camenisch2005}
{\sc Camenisch, J., and Lysyanskaya, A.}
\newblock A formal treatment of onion routing.
\newblock In {\em Annual International Cryptology Conference\/} (2005),
  Springer, pp.~169--187.

\bibitem{chaum81}
{\sc Chaum, D.}
\newblock Untraceable electronic mail, return addresses, and digital
  pseudonyms.
\newblock {\em Communications of the ACM 4}, 2 (1981), 84--88.

\bibitem{chaum1988dining}
{\sc Chaum, D.}
\newblock The dining cryptographers problem: Unconditional sender and recipient
  untraceability.
\newblock {\em Journal of Cryptology 1}, 1 (1988), 65--75.

\bibitem{cMix-patent}
{\sc Chaum, D.}
\newblock Precomputed and transactional mixing.
\newblock United States Patent, No. 10375042, Aug. 6, 2019.

\bibitem{chaum2016cmix}
{\sc Chaum, D., Das, D., Javani, F., Kate, A., Krasnova, A., De~Ruiter, J., and
  Sherman, A.~T.}
\newblock {cMix}: Anonymization by high-performance scalable mixing.
\newblock In {\em Cryptology ePrint Archive, Report 2016/008 (2016).\/} (2016).
\newblock \url{https://eprint.iacr.org/2016/008.pdf}.

\bibitem{chaum2017cmix}
{\sc Chaum, D., Das, D., Javani, F., Kate, A., Krasnova, A., De~Ruiter, J., and
  Sherman, A.~T.}
\newblock {cMix}: Mixing with minimal real-time asymmetric cryptographic
  operations.
\newblock In {\em International Conference on Applied Cryptography and Network
  Security\/} (2017), Springer, pp.~557--578.

\bibitem{CHaumUDM}
{\sc Chaum, D., Mario, Sherman, A., and Joeri}.
\newblock Udm: User discovery with minimal information disclosure.
\newblock {A}rxiv, accepted to {\it Cryptologia}, 2020.

\bibitem{chen2015}
{\sc Chen, C., Asoni, D.~E., Barrera, D., Danezis, G., and Perrig, A.}
\newblock Hornet: High-speed onion routing at the network layer.
\newblock In {\em Proceedings of the 22nd ACM SIGSAC Conference on Computer and
  Communications Security\/} (2015), pp.~1441--1454.

\bibitem{chor95pir}
{\sc Chor, B., Goldreich, O., Kushilevitz, E., and Sudan, M.}
\newblock Private information retrieval.
\newblock In {\em Proceedings of IEEE 36th Annual Foundations of Computer
  Science\/} (1995), IEEE, pp.~41--50.

\bibitem{chou2015simplest}
{\sc Chou, T., and Orlandi, C.}
\newblock The simplest protocol for oblivious transfer.
\newblock In {\em International Conference on Cryptology and Information
  Security in Latin America\/} (2015), Springer, pp.~40--58.

\bibitem{crepeau1986}
{\sc Cr{\'e}peau, C.}
\newblock A zero-knowledge poker protocol that achieves confidentiality of the
  players’ strategy or how to achieve an electronic poker face.
\newblock In {\em Conference on the Theory and Application of Cryptographic
  Techniques\/} (1986), Springer, pp.~239--247.

\bibitem{crepeau1995committed}
{\sc Cr{\'e}peau, C., van~de Graaf, J., and Tapp, A.}
\newblock Committed oblivious transfer and private multi-party computation.
\newblock In {\em Annual International Cryptology Conference\/} (1995),
  Springer, pp.~110--123.

\bibitem{danezis2004statistical}
{\sc Danezis, G., and Serjantov, A.}
\newblock Statistical disclosure or intersection attacks on anonymity systems.
\newblock In {\em International Workshop on Information Hiding\/} (2004),
  Springer, pp.~293--308.

\bibitem{desantis1995}
{\sc DeSantis, A., DiCrescenzo, G., and Persiano, G.}
\newblock Zero-knowledge arguments and public-key cryptography.
\newblock {\em Information and Computation 121}, 1 (1995), 23--40.

\bibitem{Dolev15}
{\sc Dolev, D., and Yao, A.}
\newblock On the security of public key protocols.
\newblock {\em IEEE Trans. Inf. Theor. 29}, 2 (Sept. 2006), 198–208.

\bibitem{dwork2014algorithmic}
{\sc Dwork, C., and Roth, A.}
\newblock The algorithmic foundations of differential privacy.
\newblock {\em Theoretical Computer Science 9}, 3-4 (2014), 211--407.

\bibitem{even85}
{\sc Even, S., Goldreich, O., and Lempel, A.}
\newblock A randomized protocol for signing contracts.
\newblock {\em Communications of the ACM 28}, 6 (1985), 637--647.

\bibitem{fagin96}
{\sc Fagin, R., Naor, M., and Winkler, P.}
\newblock Comparing information without leaking it.
\newblock {\em Communications of the ACM 39}, 5 (1996), 77--85.

\bibitem{furukawa2001efficient}
{\sc Furukawa, J., and Sako, K.}
\newblock An efficient scheme for proving a shuffle.
\newblock In {\em Annual International Cryptology Conference\/} (2001),
  Springer, pp.~368--387.

\bibitem{goldwasser1990}
{\sc Goldwasser, S., and Levin, L.}
\newblock Fair computation of general functions in presence of immoral
  majority.
\newblock In {\em Conference on the Theory and Application of Cryptography\/}
  (1990), Springer, pp.~77--93.

\bibitem{golle04}
{\sc Golle, P., Jakobsson, M., Juels, A., and Syverson, P.}
\newblock Universal re-encryption for mixnets.
\newblock In {\em Cryptographers’ Track at the RSA Conference\/} (2004),
  Springer, pp.~163--178.

\bibitem{halevi1996}
{\sc Halevi, S., and Micali, S.}
\newblock Practical and provably-secure commitment schemes from collision-free
  hashing.
\newblock In {\em Annual International Cryptology Conference\/} (1996),
  Springer, pp.~201--215.

\bibitem{impagliazzo1989limits}
{\sc Impagliazzo, R., and Rudich, S.}
\newblock Limits on the provable consequences of one-way permutations.
\newblock In {\em Proceedings of the Twenty-First Annual ACM Symposium on
  Theory of Computing\/} (New York, NY, USA, 1989), STOC ’89, Association for
  Computing Machinery, p.~44–61.

\bibitem{ishai2003extending}
{\sc Ishai, Y., Kilian, J., Nissim, K., and Petrank, E.}
\newblock Extending oblivious transfers efficiently.
\newblock In {\em Annual International Cryptology Conference\/} (2003),
  Springer, pp.~145--161.

\bibitem{jakobsson1999flash}
{\sc Jakobsson, M.}
\newblock Flash mixing.
\newblock In {\em Proceedings of the Eighteenth Annual ACM Symposium on
  Principles of Distributed Computing\/} (1999), pp.~83--89.

\bibitem{jakobsson2001}
{\sc Jakobsson, M., and Juels, A.}
\newblock An optimally robust hybrid mix network.
\newblock In {\em Proceedings of the Twentieth Annual ACM Symposium on
  Principles of Distributed Computing\/} (2001), pp.~284--292.

\bibitem{javani2020bvot}
{\sc Javani, F., and Sherman, A.~T.}
\newblock {BVOT}: Self-tallying boardroom voting with oblivious transfer.
\newblock {\em arXiv preprint arXiv:2010.02421\/} (2020).
\newblock \url{https://arxiv.org/abs/2010.02421}.

\bibitem{kedogan2002limits}
{\sc Kedogan, D., Agrawal, D., and Penz, S.}
\newblock Limits of anonymity in open environments.
\newblock In {\em International Workshop on Information Hiding\/} (2002),
  Springer, pp.~53--69.

\bibitem{kesdogan1998}
{\sc Kesdogan, D., Egner, J., and B{\"u}schkes, R.}
\newblock Stop-and-go-mixes providing probabilistic anonymity in an open
  system.
\newblock In {\em International Workshop on Information Hiding\/} (1998),
  Springer, pp.~83--98.

\bibitem{kil88}
{\sc Kilian, J.}
\newblock Founding crytpography on oblivious transfer.
\newblock In {\em Proceedings of the Twentieth Annual ACM Symposium on Theory
  of Computing\/} (1988), pp.~20--31.

\bibitem{krawczyk2010hkdf}
{\sc Krawczyk, H.}
\newblock Cryptographic extraction and key derivation: The {HKDF} scheme.
\newblock In {\em Annual Cryptology Conference\/} (2010), Springer,
  pp.~631--648.

\bibitem{HKDF}
{\sc Krawczyk, H., and Eronen, P.}
\newblock {MAC}-based extract-and-expand key derivation function {(HKDF)}.
\newblock Internet Engineering Task Force {(IETF)}, Request for Comments: 5869,
  May 2010.
\newblock \url{https://tools.ietf.org/html/rfc5869}.

\bibitem{kwon2016}
{\sc Kwon, A., Lazar, D., Devadas, S., and Ford, B.}
\newblock Riffle: An efficient communication system with strong anonymity.
\newblock {\em Proceedings on Privacy Enhancing Technologies 2016}, 2 (2016),
  115--134.

\bibitem{moller2003}
{\sc M{\"o}ller, B.}
\newblock Provably secure public-key encryption for length-preserving chaumian
  mixes.
\newblock In {\em Cryptographers’ Track at the RSA Conference\/} (2003),
  Springer, pp.~244--262.

\bibitem{Naor01}
{\sc Naor, M., and Pinkas, B.}
\newblock Efficient oblivious transfer protocols.
\newblock In {\em 12th Annual ACM-SIAM Symposium on Discrete Algorithms\/}
  (2001), pp.~448--457.

\bibitem{naor2005compsec}
{\sc Naor, M., and Pinkas, B.}
\newblock Computationally secure oblivious transfer.
\newblock {\em Journal of Cryptology 18}, 1 (2005), 1--35.

\bibitem{nielsen2012new}
{\sc Nielsen, J.~B., Nordholt, P.~S., Orlandi, C., and Burra, S.~S.}
\newblock A new approach to practical active-secure two-party computation.
\newblock In {\em Annual Cryptology Conference\/} (2012), Springer,
  pp.~681--700.

\bibitem{nurmi91}
{\sc Nurmi, H., Salomaa, A., and Santean, L.}
\newblock Secret ballot elections in computer networks.
\newblock {\em Computers \& Security 10}, 6 (1991), 553--560.

\bibitem{ohkubo2000l}
{\sc Ohkubo, M., and Abe, M.}
\newblock A length-invariant hybrid mix.
\newblock In {\em International Conference on the Theory and Application of
  Cryptology and Information Security\/} (2000), Springer, pp.~178--191.

\bibitem{overlier2007}
{\sc {\O}verlier, L., and Syverson, P.}
\newblock Improving efficiency and simplicity of tor circuit establishment and
  hidden services.
\newblock In {\em International Workshop on Privacy Enhancing Technologies\/}
  (2007), Springer, pp.~134--152.

\bibitem{park93}
{\sc Park, C., Itoh, K., and Kurosawa, K.}
\newblock Efficient anonymous channel and all/nothing election scheme.
\newblock In {\em Workshop on the Theory and Application of of Cryptographic
  Techniques\/} (1993), Springer, pp.~248--259.

\bibitem{peikert2008framework}
{\sc Peikert, C., Vaikuntanathan, V., and Waters, B.}
\newblock A framework for efficient and composable oblivious transfer.
\newblock In {\em Annual International Cryptology Conference\/} (2008),
  Springer, pp.~554--571.

\bibitem{pfitzmann2001}
{\sc Pfitzmann, A., and K{\"o}hntopp, M.}
\newblock Anonymity, unobservability, and pseudonymity—a proposal for
  terminology.
\newblock In {\em Designing Privacy Enhancing Technologies\/} (2001), Springer,
  pp.~1--9.

\bibitem{pfitzmann85}
{\sc Pfitzmann, A., and Waidner, M.}
\newblock Networks without user observability—design options.
\newblock In {\em Workshop on the Theory and Application of of Cryptographic
  Techniques\/} (1985), Springer, pp.~245--253.

\bibitem{piotrowska2017loopix}
{\sc Piotrowska, A.~M., Hayes, J., Elahi, T., Meiser, S., and Danezis, G.}
\newblock The {Loopix} anonymity system.
\newblock In {\em 26th {USENIX} Security Symposium ({USENIX} Security 17)\/}
  (2017), pp.~1199--1216.

\bibitem{Rabin81}
{\sc Rabin, M.~O.}
\newblock How to exchange secrets with oblivious transfer.
\newblock {\em Technical Report TR-81, Aiken Computation Lab, Harvard
  University\/} (1981).

\bibitem{rackoff1993cryptographic}
{\sc Rackoff, C., and Simon, D.~R.}
\newblock Cryptographic defense against traffic analysis.
\newblock In {\em Proceedings of the twenty-fifth annual ACM symposium on
  Theory of computing\/} (1993), pp.~672--681.

\bibitem{raymond2001traffic}
{\sc Raymond, J.-F.}
\newblock Traffic analysis: Protocols, attacks, design issues, and open
  problems.
\newblock In {\em Designing Privacy Enhancing Technologies\/} (2001), Springer,
  pp.~10--29.

\bibitem{reed98}
{\sc Reed, M., Syverson, P., and Goldschlag, D.}
\newblock {Anonymous Connections and Onion Routing}.
\newblock {\em IEEE J-SAC 16}, 4 (1998), 482--494.

\bibitem{serjantov2002towards}
{\sc Serjantov, A., and Danezis, G.}
\newblock Towards an information theoretic metric for anonymity.
\newblock In {\em International Workshop on Privacy Enhancing Technologies\/}
  (2002), Springer, pp.~41--53.

\bibitem{serjantov2002}
{\sc Serjantov, A., Dingledine, R., and Syverson, P.}
\newblock From a trickle to a flood: Active attacks on several mix types.
\newblock In {\em International Workshop on Information Hiding\/} (2002),
  Springer, pp.~36--52.

\bibitem{SRP2020}
{\sc Sherman, A.~T., Lanus, E., Liskov, M., Zieglar, E., Chang, R.,
  Golaszewski, E., Wnuk-Fink, R., Bonyadi, C.~J., Yaksetig, M., and Blumenfeld,
  I.}
\newblock Formal methods analysis of the secure remote password protocol.
\newblock In {\em Logic, Language, and Security: {E}ssays dedicated to {A}ndre
  {S}cedrov on the occassion of his 65th birthday\/} (February 2020), e.~a.
  Nigam, Ed., vol.~12300 of {\em LNCS Festscrift}, Springer, pp.~103--126.
\newblock Available as \url{https://arxiv.org/pdf/2003.07421.pdf}.

\bibitem{syta2014security}
{\sc Syta, E., Corrigan-Gibbs, H., Weng, S.-C., Wolinsky, D., Ford, B., and
  Johnson, A.}
\newblock Security analysis of accountable anonymity in dissent.
\newblock {\em ACM Transactions on Information and System Security (TISSEC)
  17}, 1 (2014), 1--35.

\bibitem{tor2004}
{\sc Syverson, P., Dingledine, R., and Mathewson, N.}
\newblock Tor: The secondgeneration onion router.
\newblock In {\em USENIX Security\/} (2004), pp.~303--320.

\bibitem{tyagi2017stadium}
{\sc Tyagi, N., Gilad, Y., Leung, D., Zaharia, M., and Zeldovich, N.}
\newblock Stadium: A distributed metadata-private messaging system.
\newblock In {\em Proceedings of the 26th Symposium on Operating Systems
  Principles\/} (2017), pp.~423--440.

\bibitem{van2015vuvuzela}
{\sc Van Den~Hooff, J., Lazar, D., Zaharia, M., and Zeldovich, N.}
\newblock Vuvuzela: Scalable private messaging resistant to traffic analysis.
\newblock In {\em Proceedings of the 25th Symposium on Operating Systems
  Principles\/} (2015), pp.~137--152.

\bibitem{wolinsky2012dissent}
{\sc Wolinsky, D.~I., Corrigan-Gibbs, H., Ford, B., and Johnson, A.}
\newblock Dissent in numbers: Making strong anonymity scale.
\newblock In {\em 10th {USENIX} Symposium on Operating Systems Design and
  Implementation ({OSDI} 12)\/} (2012), pp.~179--182.

\end{thebibliography}

\appendix

\mysavetinyspace
\section{Appendix}

\mysavespace
\subsection{Notation}  
\label{sec:notation}

\begin{table}[h!]
\centering
\caption{Notation.}
\label{tab:notation}
\begin{tabular*}{\columnwidth}{r|l} 
$C$             &   container of nodes at Level~1    \\ 
$\Psi$          &   batch of messages at Level-2 nodes \\
$\Phi$          &   bucket of messages at Level-3 nodes \\
$\beta_1$       &	batch size at Level-1 nodes \\
$\beta_2$       &	batch size of real messages at Level-2 nodes \\
$\gamma$        &	number of messages in pub.\ lists of Level-3 nodes \\
$\zeta$         &   number of messages from which to choose in OT\\
$\rho$          &	number of passive Level-3 nodes\\
$\alpha$        &	number of active Level-3 nodes\\
$\sigma_{A,B}$  &	shared secret between sender $A$ and receiver $B$ \\
$k_{A_j,B_j}$   &	tag of message $j$ from sender $A$ to receiver $B$ \\
$P$             &	partition of the set of $\beta_2$ indexes \\
$x_j$           &	payload of message $j$ \\
$E[k,M]$        &	encryption of message $M$ under key $k$ \\
$p_b$           &	public key of $B$ \\
$s_b$           &   secret key of $B$\\
$h$             &	hash function \\
$\lambda$       &   number of batches in each message repository\\
$\mathcal{R}_i$ &   round $i$\\
${\omega}$      &   residual pool size\\
$\Omega$        &   number of incoming messages into pool \\
$T_1$           &   maximum interval for verifying messages \\
$T_2$           &   maximum interval for dummy requests \\
$n$             &   nonce\\
$ts$            &   timestamp\\
$\xi$           &   counter deviation bound\\
\end{tabular*}

\end{table} 

\mysavespace
\subsection{Proof of Theorem~1} 
\label{sec:proof}

One way to reason about the anonymity properties of AOT is in terms of a game between a {\it challenger} $\mathcal{C}$ and an {\it adversary} 
$\mathcal{A}$ \cite{brickell2006efficient,syta2014security,chaum2016cmix}. 
The challenger $\mathcal{C}$ performs the protocol on behalf of honest mixnodes and users, while adversary $\mathcal{A}$---a {\it probabilistic polynomial time (PPT)} Turing machine---performs the protocol on behalf of compromised mixnodes and users.

The covert adversary can compromise all of the mixnodes and a fraction of the users. The adversary can observe the internal states of all compromised mixnodes, and the adversary may inject, drop, or delay messages. The adversary can also eavesdrop on communications between mixnodes and users.


The game between the challenger and the adversary works as follows:

\begin{enumerate}

\item[\textbf{Step 1.}] The challenger $\mathcal{C}$ runs AOT for all of the honest users, and shares all public information with the adversary.

\item[\textbf{Step 2.}] For as many times as the adversary $\mathcal{A}$ wants, $\mathcal{C}$ simulates the honest users, and $\mathcal{A}$ runs the protocol with $\beta_2$ messages with inputs from $\mathcal{C}$.

\item[\textbf{Step 3.}] The adversary chooses two honest users $c_0$ and $c_1$ and a message $m$. The challenger chooses a bit $b$ at random by tossing a uniform random coin. The challenger sets $c_b$ to be the receiver of the message $m$, sets $c_{1-b}$ to be a user who does not have a message in the publication list, and initiates a dummy request with AOT. Let $m^\prime$
be the message in the publication list that $c_{1-b}$ pretends to ask for. $\mathcal{C}$ runs the protocol with $c_b$ and $c_{1-b}$ and sends $m^\prime$ to $\mathcal{A}$.

\item[\textbf{Step 4.}] After the challenge phase, the adversary runs the AOT protocol with $\beta_2$ messages as many times as it desires to.

\item[\textbf{Step 5.}] The adversary outputs its guess for $b$.

\end{enumerate}  

Let $\langle\mathcal{A}|\mathcal{C}\rangle$ denote the adversary's output in the game. The adversary's advantage in the anonymity game is\linebreak
{$|Pr[\langle\mathcal{A}|\mathcal{C}\rangle=b]-1/2|$.}

\begin{definition}\label{def:rec-anon}
A protocol maintains receiver anonymity if the adversary's advantage in the anonymity game is negligible.
\end{definition}

\begin{proof} (of Theorem~\ref{thm:rac})

We reduce the receiver security of OT to the receiver anonymity of AOT.

Define the {\it OTRS Problem} as follows: in an $\textrm{OT}^\eta_1$ oblivious transfer session with the sender $\mathcal{A}_{OT}$ with $\eta$ strings $s_1, s_2, \dots s_\eta$, given $1 \leq i < j \leq \eta$ where the receiver chooses either $s_i$ or $s_j$, distinguish whether the receiver chose $s_i$ or $s_j$.

We define the OTRS game between the challenger $\mathcal{C}_{OT}$ and the adversary $\mathcal{A}_{OT}$ as follows: 

\begin{enumerate}

    \item[\textbf{Step 1.}] $\mathcal{A}_{OT}$ runs the protocol for the sender, and $\mathcal{C}_{OT}$ runs the protocol for the receiver.
    
    \item[\textbf{Step 2.}] For as many times as $\mathcal{A}_{OT}$ wants to, $\mathcal{C}_{OT}$ engages in $\textrm{OT}^\eta_1$ with $\mathcal{A}_{OT}$.
    
    \item[\textbf{Step 3.}] $\mathcal{A}_{OT}$ generates the strings $s_1, s_2, \dots s_\eta$. $\mathcal{C}_{OT}$ chooses a string with $\textrm{OT}^\eta_1$; let $s_a$ denote $\mathcal{C}_{OT}$'s choice. $\mathcal{C}_{OT}$ sends the indexes $i$ and $j$, for all $1\leq i < j \leq \eta$, to $\mathcal{A}_{OT}$ where $a\in\{i,j\}$.
    
    \item[\textbf{Step 4.}] For as many times as $\mathcal{A}_{OT}$ wants to, $\mathcal{C}_{OT}$ engages in $\textrm{OT}^\eta_1$ with $\mathcal{A}_{OT}$.
    
    \item[\textbf{Step 5.}] $\mathcal{A}_{OT}$ outputs  its guess of $a$.
    
    
\end{enumerate}

Any instance of the OTRS game can be transformed into an instance of the anonymity game as follows.

In Step~1 of the anonymity game, $\mathcal{C}$ will initiate an OTRS game with the challenger $\mathcal{C}_{OT}$.
In Step~2 of the anonymity game, $\mathcal{C}$ simulates the honest users as many times as $\mathcal{A}$ desires.

In Step~3 of the anonymity game, after $\mathcal{C}$ receives the target message $m$ and two honest users $c_0$ and $c_1$ from $\mathcal{A}$, it chooses $c_b$ (by tossing a coin) to be the recipient of $m$, and $c_{1-b}$ to be user that initiates a dummy request to the publication list. $\mathcal{C}$ will set the message that $\textrm{OT}^{\beta_2}_1$ will pretend to ask for in the dummy request to be the choice of $\mathcal{C}_{OT}$ in the oblivious transfer in the OTRS game. $\mathcal{C}$ will run the protocol for $c_b$ and $c_{1-b}$ as follows.

In $\mathcal{C}$'s simulation of $c_b$, user $c_b$ will ask for $m$ with $\textrm{OT}^{\beta_2}_1$ from $\mathcal{A}$. To simulate $c_{1-b}$, challenger $\mathcal{C}$ starts Step~3 of the OTRS game with $\mathcal{C}_{OT}$, where $\mathcal{C}$'s strings are the messages in the publication list of the Level-3 node. During the $\textrm{OT}^{\beta_2}_1$ oblivious transfer session of the OTRC game, when $\mathcal{C}$ receives a message from $\mathcal{C}_{OT}$, it will send the message to $\mathcal{A}$, pretending that the message originated from $c_{1-b}$. When $c_{1-b}$ receives a message from $\mathcal{A}$, $\mathcal{C}$ will forward the message to $\mathcal{C}_{OT}$, pretending that the message originated from $\mathcal{C}$.

{At the end of Step~3 of the OTRS game, $\mathcal{C}_{OT}$ sends the indexes $i$ and $j$ to $\mathcal{C}$. Let $m_i$ and $m_j$ denote the messages corresponding to $i$ and $j$,
respectively, from the publication repository, and let $m_a$ denote $\mathcal{C}_{OT}$'s choice. At the end of Step~3 of the anonymity game---as its guess for $m_a$---$\mathcal{C}$ sends $m_i$ to $\mathcal{A}$. 
If $\mathcal{A}$ guesses the value $b$ correctly, the challenger $\mathcal{C}$ knows that its guess for $m_a$ is correct and $\mathcal{A}$ has concluded that $c_{1-b}$ has asked for $m_i$ in a dummy request. Therefore, in this case,
$\mathcal{C}$ outputs $i$ for its guess of $a$ to $\mathcal{C}_{OT}$. 
If $\mathcal{A}$ guesses the value $b$ incorrectly, then $\mathcal{C}$ outputs $j$ for its guess of $a$.}

If $\mathcal{A}$'s advantage in the anonymity game is non-negligible, then $\mathcal{A}_{OT}$ can distinguish between $s_i$ or $s_j$,
contradicting the fact that AOT uses an OT protocol that provides receiver security.

\end{proof}

\mysavespace
\subsection{Anonymity of a Pool-Mix AOT} \label{sec:pool}

Using entropy, we analyze the anonymity of a pool-mix version
of AOT.  Pool-mixes offer additional defenses to blending attacks.

A {\it pool mix}~\cite{serjantov2002} stores a residual pool of $\omega$ messages (similar to AOT's message repository); whenever it receives $\Omega$ messages, it adds them into the pool, selects $\Omega$ messages randomly from the $\omega+\Omega$ messages in the pool and outputs them. The random selection of messages causes some of the messages to be delayed for a long time---potentially infinitely. 
By contrast, in AOT, Level-3 nodes store messages before publication for less than $\tau$ seconds; that is, messages are chosen from the message repository from the last $\lambda$ batches such that each message is published in less than $\tau$ time after it enters a Level-3 node. 

If this maximum delay is removed from the system, AOT can be converted into a pool mix as follows: AOT selects $\beta_2/\alpha$ messages randomly from $\beta_2(\lambda+1)/(2\alpha)$ messages in the message repository---as opposed to selecting $\beta_2/(\alpha \lambda)$ from each bucket---for each publication (see Figure~\ref{fig:repo}).

When AOT selects messages randomly from each bucket in the message repository, AOT publishes messages from each of the last $\lambda$ rounds with the same probability. Therefore, the size of the anonymity set is an appropriate measure of anonymity. 
Serjantov and Danezis~\cite{serjantov2002towards}, however, show that anonymity set is not an appropriate measure to analyze the anonymity of pool mixes, because size of the anonymity set does not reflect  that messages from different rounds have different probabilities of exiting the mix at each round. Instead,
Serjantov and Danezis propose using entropy to measure the anonymity of pool mixes. 

Serjantov and Danezis define the effective size of the sender
anonymity set as $2^E$, where $E$ is the entropy of the probability distribution $\mathcal{U}$, and $\mathcal{U}$ is ``the attacker's a-posteriori probability distribution'' of all the users being the sender of a specific message.  
The same definition applies for the receiver anonymity set.

Applying the analysis of Serjantov and Danezis, for a pool mix with pool size $\omega$, where $\Omega$ messages are output at each round, when the number of rounds $k$ approaches infinity, we have:

\begin{equation}
    \lim_{k\rightarrow \infty} E =  \left(1+ \frac{\omega }{ \Omega} \right)\log\left( \omega+\Omega \right) - \frac{ \omega }{ \Omega}\log {\omega}     .  
\end{equation}

Considering the message repository of each Level-3 node in AOT, to compare the anonymity sets of AOT with those of a pool-mix version of AOT, we set $\omega = \beta_2(\lambda-1)/(2\alpha)$ and $\Omega = \beta_2/\alpha$. For each Level-3 node we have:

\begin{equation}\label{ent-aot}
    \lim_{k\rightarrow \infty} E = \log \frac{\lambda\beta_2}{\alpha}  + \log \frac{(\lambda+1)^\frac{\lambda+1}{2}}{2\lambda(\lambda-1)^\frac{\lambda-1}{2}},
\end{equation}

\noindent where
$\lambda\beta_2/\alpha$ is the size of the anonymity set of each Level-3 node in AOT. 

Therefore, removing the maximum delay from AOT and converting AOT to a pool mix will add

\begin{equation}
 \log \frac{(\lambda+1)^\frac{\lambda+1}{2}}{2\lambda(\lambda-1)^\frac{\lambda-1}{2}}   
\end{equation}

\noindent to the entropy of the probability distribution $\mathcal{U}$. That is, in a pool-mix version of AOT, the sender anonymity set size will increase by a factor of

\begin{equation}
2^ {\left(\log \frac{(\lambda+1)^\frac{\lambda+1}{2}}{2\lambda(\lambda-1)^\frac{\lambda-1}{2}}  \right) }.    
\end{equation}

A poll-mix version of AOT will result in the following latency:
messages will be published after an average of $(\lambda+1)/2$ rounds, with a variance of $(\lambda+1)^2(\lambda-1)/8$ rounds. By contrast, regular AOT publishes messages after an average of $\lambda/2$ rounds, with a variance of $(\lambda^2-1)/12$ rounds.

For example, if we set $\lambda=9$, for each Level-3 node,
the size of the effective anonymity set of a pool-mix version of AOT will be larger than the size of anonymity set of AOT by a factor of $2^{0.44}\approx 1.35$---at the expense of potentially delaying some messages infinitely. It will take the pool version of AOT $5$ rounds on average to deliver a message with a variance of $100$ rounds.


\mysavespace
\subsection{{Public-Key Encryption with Ciphertext Integrity}}
\label{Appendix-A}

AOT requires
the encryption function $E$ (Section~\ref{sec:AOT}) to provide 
confidentiality and integrity. 
For example, AOT could use
Bernstein's~\cite{crypto_box} 
crypto\_box function, which
performs {public-key encryption }
as follows.

\begin{enumerate}
    \item Generate a symmetric key at random.
    \item Encrypt the packet using the symmetric key.\label{it:a}
    \item Hash the encrypted packet using a cryptographic hash function.
    \item Sign the hash value using Alice’s secret key.
    \item Encrypt the symmetric key, hash, and signature using Bob’s public key.\label{it:b}
    \item Concatenate 
    the ciphertext (from Step~\ref{it:b}) with the encrypted packet (Step~\ref{it:a}).
\end{enumerate}

\mysavespace
\subsection{Acronyms and Abbreviations}
\label{Appendix-B}

\bigskip

\begin{tabular}{ll}
ACS & Anonymous Comumnication System\\
AOT & Anonymity by Oblivious Transfer\\
GPA & Global Passive Adversary\\
OT & Oblivious Transfer\\
MAC & Message Authentication Code\\
PIR & Private Information Retrieval\\
PPT & Probabilistic Polynomial-Time Turning Machine\\
RAC & Receiver Anonymity with Compromised network \\
\end{tabular}


\end{document}

\end{document}